\documentclass[reprint,aps,prx,floatfix,notitlepage,superscriptaddress]{revtex4-2}

\usepackage{graphicx}
\usepackage{amsmath}
\usepackage{amssymb}
\usepackage{bm}
\usepackage[dvipsnames]{xcolor}
\usepackage[colorlinks=true,allcolors=Green]{hyperref}

\usepackage[separate-uncertainty=true, multi-part-units=single, parse-numbers=false]{siunitx}
\DeclareSIUnit{\mol}{mol}
\DeclareSIUnit{\Molar}{M}
\DeclareSIUnit{\Kelvin}{K}

\newcommand{\pc}[1]{({#1})}
\newcommand{\p}[1]{#1}

\newcommand*\dif{\mathop{}\!\mathrm{d}}
\renewcommand{\vec}[1]{{\bm #1}}

\renewcommand\O{{\textsc{o}}}
\newcommand\A{{\textsc{a}}}
\newcommand\B{{\textsc{b}}}

\newcommand\Dc{D_\text{c}}
\newcommand\Dp{D_\text{p}}
\newcommand\cutoff{\ell}
\newcommand\Rbdr{R_\text{bdr}}
\newcommand{\sst}[1]{{#1}^{\text{ss}}}

\begin{document}

\title{Emergent single-species non-reciprocity from bistable chemical dynamics}

\author{Jakob Metson}
\affiliation{Max Planck Institute for Dynamics and Self-Organization (MPI-DS), 37077 G\"ottingen, Germany}

\author{Ramin Golestanian}
\affiliation{Max Planck Institute for Dynamics and Self-Organization (MPI-DS), 37077 G\"ottingen, Germany}
\affiliation{Rudolf Peierls Centre for Theoretical Physics, University of Oxford, Oxford OX1 3PU, United Kingdom}

\date{\today}

\begin{abstract}
\noindent
The appearance of emergent symmetries in complex systems with components that can form composite units provides us with opportunities for design and control of exotic phase behaviour, for example by exploiting the dynamical symmetry breaking associated with them. We present a novel mechanism for the emergence of non-reciprocal interactions in a single-species suspension of chemically active colloids made out of semi-permeable vesicles, which encapsulate enzymes that catalyze a non-linear chemical reaction. Bistable chemical dynamics enables the colloidal reaction chamber to act as a net producer or consumer of a chemical, depending on the selected values of the chemical concentrations inside and around it. Since the internal chemical state of the colloid depends on the dynamic chemical concentrations rather than the material parameters, two identically produced colloids can present different effective chemical interactions within the same system upon responding to the corresponding gradients via diffusiophoresis. Furthermore, the colloids can spontaneously and reversibly switch between being effective consumers or producers. As a consequence, the colloids can dynamically switch between ignoring, attracting, repelling, and chasing each other, in a non-reciprocal manner. This flexibility can be exploited by manipulation of tuning parameters to induce bifurcations in the chemical dynamics, resulting in a robust control over the interaction motifs, and rich emergent dynamics such as spontaneous many-body polar swarming.
\end{abstract}

\maketitle

\section{Introduction}

Achieving non-equilibrium function requires time-reversal symmetry breaking, which can be afforded as per Curie principle via breaking of other forms of symmetry, such as parity and time-translation \cite{Rousselet1994,Pumm2022}. While evolution has enabled biological systems to achieve spontaneous symmetry breaking \cite{Kruse2004,Guirao2007}, it is important to investigate how such capability can be engineered in artificial systems \cite{Chen2025}. Rare examples of such characteristics include spontaneous chiral symmetry breaking of elastic filaments that leads to active motion \cite{Shi2022,Butler2026}, 
proto-cell division of metabolically active droplets due to Rayleigh instability \cite{Zwicker2016,Golestanian2016nphys},
and emergence of discrete symmetry in the form of run-and-tumble behaviour \cite{Polin2009,Bennett2013,Soto2015PRE} and gating activity \cite{Bonthuis2014} in systems with continuos degrees of freedom due to mechanical constraints. While in these examples the emergence of symmetry breaking has been the result of serendipity, it will be desirable to develop conceptual routes to achieving such level of complex functionality that parallels what is observed in living systems. 

A promising medium to explore towards this end is non-reciprocal active matter, in which effective interactions emerge that seemingly break Newton's third law of physics \cite{Soto2014PRL,Ivlev2015PRX}, namely, the mechanical response of particle $\A$ to particle $\B$ is not equal and opposite to the response of particle $\B$ to particle $\A$. Non-reciprocal interactions between microscopic particles, which naturally break time-reversal symmetry, have been shown to lead to many interesting collective phenomena \cite{Saha2020,You2020PNAS,Fruchart2021N,Kreienkamp2022NJP,Dinelli2023NC}, including shape-shifting multifarious self-organization \cite{Osat2022} and fast and efficient self-organization of metabolic cycles that may have been relevant in early forms of life \cite{OuazanReboul2023b}.

For chemically active colloids that exhibit non-reciprocity via diffusiophoretic response, interactions are analogous to generalized electrostatic or gravitational forces, in which the charge or mass that produces the field can be different from the charge or mass that responds to the field generated by other particles \cite{Soto2014PRL,Agudo-Canalejo2019PRL}. While this symmetry breaking leads to rich phase behaviour, it would be remarkable if the acquisition of charge or mass can be enabled through spontaneous symmetry breaking, in analogy with the Higgs mechanism, rather than being externally hardwired in the system, for example by using two populations of Janus particles \cite{Saha2019NJP}, 
asymmetrically shaped particles \cite{Gupta2022PRE}, or externally controlled vision cones \cite{Lavergne2019S}. 

\begin{figure}[!ht]
\centering
\includegraphics[width=\linewidth]{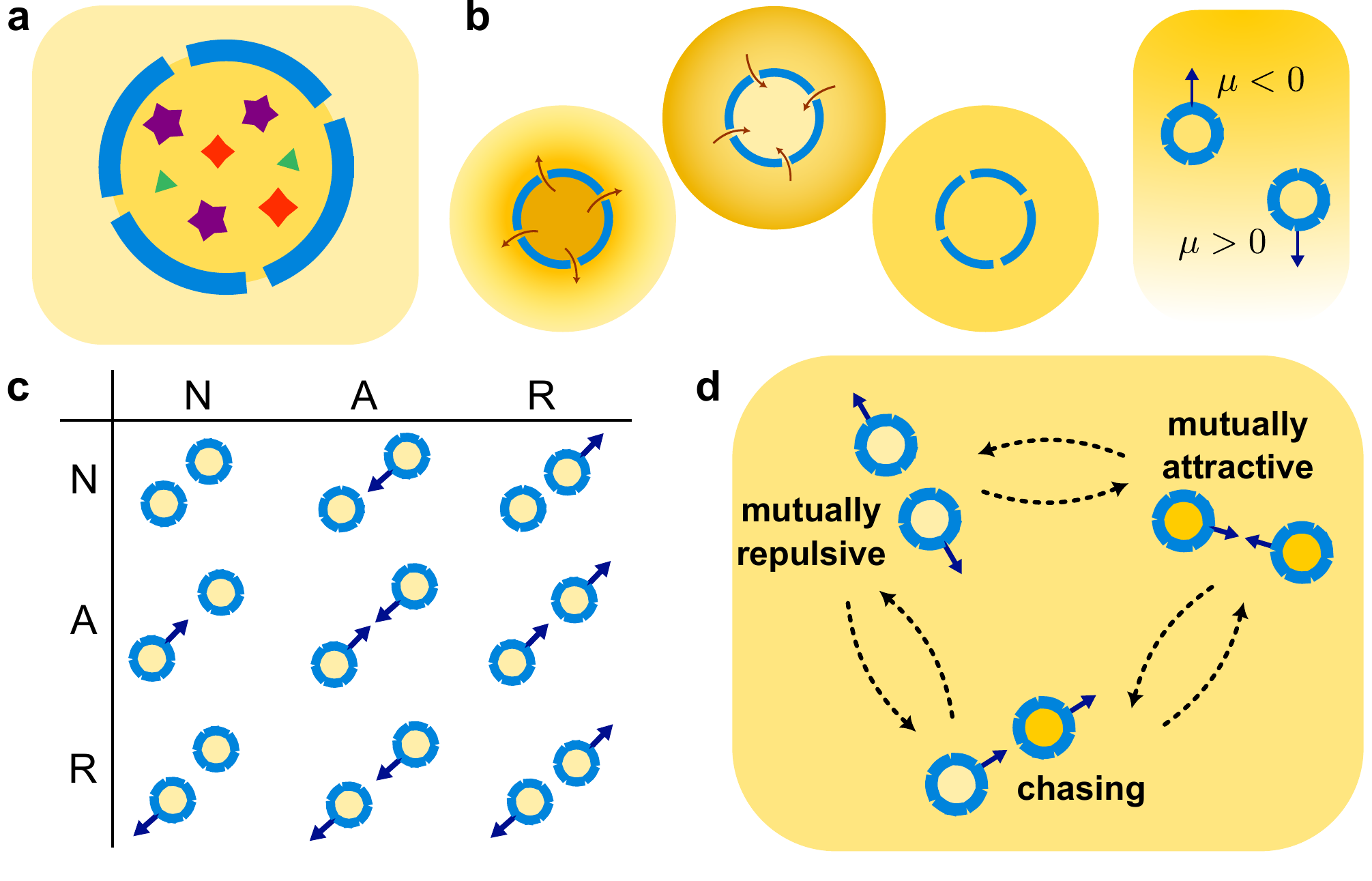}
\caption{\textbf{Mechanochemically active colloids.}
\pc{a} A colloidal reaction chamber is formed from a semi-permeable vesicle that encapsulates enzymes, which catalyze a non-linear reaction involving a representative chemical (depicted by continuous yellow shading).
\pc{b} A colloid with an internal concentration higher (lower) than the external concentration is an effective producer (consumer) of the chemical. This leads to gradients in the external chemical concentration around the colloids. Alternatively, when the internal chemical concentration matches that of the surroundings, there will be no gradients. In response to chemical concentration gradients, the colloids move due to phoretic response, with the direction being controlled by the sign of the corresponding mobility $\mu$.
\pc{c} The interplay between chemical activity and diffusiophoresis leads to a rich variety of interaction motifs between the colloids, including non-reciprocal chasing interactions. Each colloid can experience neutral (N), attractive (A), or repulsive (R) interactions towards other colloids.
\pc{d} Identically prepared colloids can exhibit a range of possible interactions depending on the corresponding internal chemical states, and dynamically switch between them.
}
\label{fig:cartoon}
\end{figure}

\section{Results}

We consider colloidal particles that are active reaction chambers in the form of semi-permeable vesicles (see Fig.~\ref{fig:cartoon}\p{a}, and Ref. \cite{BorgesFernandes2025} for an example). Their activity manifests in two distinct ways: they {\em chemically} interact with a representative chemical in the medium and {\em mechanically} respond to chemical gradients produced by others (Fig.~\ref{fig:cartoon}\p{b}). The vesicles encapsulate enzymes (e.g. belonging to a reaction network) that catalyze a non-linear reaction designed to enable the existence of multiple distinct internal chemical states. The resulting dynamically generated gradient by each particle can initiate diffusiophoretic response of other particles, thereby introducing a novel class of phoretic active matter with active internal chemical dynamics \cite{Golestanian2022a}. The multiplicity of the internal states naturally lends itself to a combinatorial space of different interaction motifs between two particles (Fig.~\ref{fig:cartoon}\p{c}), which includes an emergent non-reciprocal chasing interaction between identically prepared particles (Fig.~\ref{fig:cartoon}\p{d}).  

Our proposal provides a rare realization of emergent polar symmetry in a system of identically prepared non-polar particles. Emergent compound particles and emergent symmetries play a key role in determining non-trivial collective behaviour in exotic condensed matter systems, such as Cooper-pairs in BCS theory of superconductivity \cite{BCS1957} and fractionalization and spin-charge separation in theories of high-temperature superconductivity \cite{LeeRMP2006}. While such complexations often ``neutralize'' features such as polarity and spin, by e.g. creating bosonic pseudo-particles that can undergo Bose condensation, in our case the particles combine to create polarity; something akin to two bosons forming a fermionic complex in this analogy, which would be unthinkable. Importantly, the emergent polar symmetry arising from the complexation can undergo spontaneous symmetry breaking, giving rise to novel phase behaviour in the system \cite{Pisegna2024}.

We start by demonstrating the bistable chemical dynamics with a single colloid coupled to the chemical background. We then show how this can lead to a range of interactions between two colloids, classifying how the different interaction possibilities change with different parameters, and how this can be used to dynamically control the behaviour of identically prepared chemically active colloids. Next, we look at the collective dynamics in suspensions of many colloids, highlighting the emergence of self-propelled chasing clusters. 
The details of the theoretical framework are presented in the Methods section.

\begin{figure*}[t]
\centering
\includegraphics[width=\linewidth]{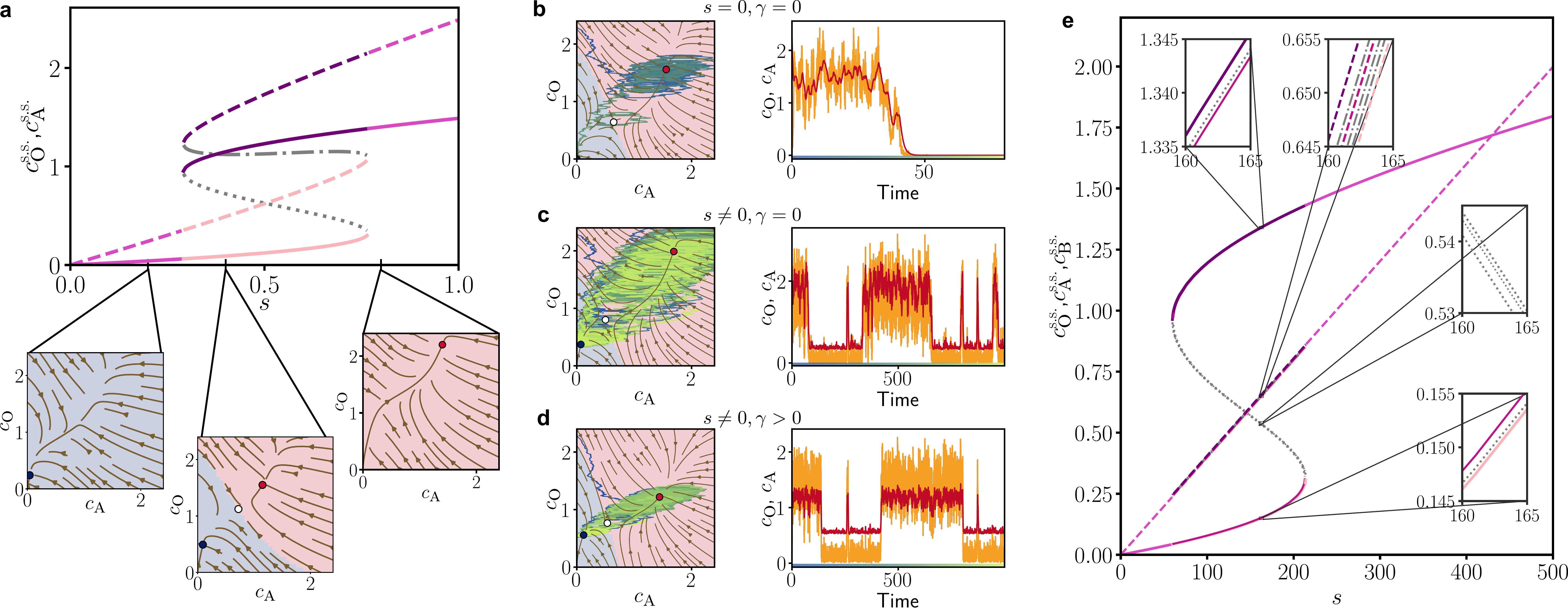}
\caption{\textbf{Space-free chemical dynamics.} 
\pc{a} Bifurcation diagram of the fixed points with a single colloid and the background, as $s$ is increased. 
Dashed and solid lines denote stable $(\sst{c}_\O,\sst{c}_\A)$ fixed points, dot-dashed and dotted lines indicate unstable $(\sst{c}_\O,\sst{c}_\A)$ fixed points.
Phase-space flows are shown for representative values of $s$, with the stable fixed points and their basins of attraction indicated, as well as any unstable fixed points.
\pc{b}--\pc{d} Chemical dynamics of a single colloid and the background. In the left panels, a stochastic trajectory of the chemical dynamics is shown overlaid on the phase-space flow. The trajectories are re-plotted in the panels on the right, with the red (darker) curves showing $c_\O(t)$ and the orange (lighter) curves showing $c_\A(t)$. The trajectories in the left panels are coloured by time, with the corresponding colouring shown along the horizontal axis of the panels on the right.
\pc{e} Bifurcation diagram of the fixed points with two colloids and the background, as $s$ is increased. 
Different colours correspond to the different stable fixed points at each $s$ value. Solid lines show stable $(\sst{c}_\A,\sst{c}_\B)$ pairs, which due to the identical nature of the colloids can be interchanged. Dashed lines show the corresponding $\sst{c}_\O$ values. Unstable fixed points are indicated by grey lines, dotted for $(\sst{c}_\A,\sst{c}_\B)$ and dot-dashed for $\sst{c}_\O$. Fixed points with $\sst{c}_\A=\sst{c}_\B$ are marked by thicker lines.
}
\label{fig:chem_dyn}
\end{figure*}

\subsection{Bistable chemical dynamics}

The semi-permeable membrane renders the chemical dynamics inside the colloids different to that in the medium. The internal chemical dynamics of a colloid is characterized by a chemical concentration $c$ and a non-linear rate function of the form $f(c) = \frac{a  K \,c^2}{K^2+c^2} - b c$, involving internal production rate $a$ and degradation rate $b$, as well as a Michaelis constant $K$ that sets the overall concentration scale of the system. The colloids exchange chemicals with the bulk surroundings through a semi-permeable membrane at rate $\Gamma$. For a given colloid $\A$ whose inner concentration $c_\A(t)$ is coupled to a background chemical concentration $c_\O(t)$, the chemical dynamics reads
\begin{equation}
\frac{\dif c_\A}{\dif t}=f(c_\A)+\Gamma \left(c_\O-c_\A\right),    
\end{equation}
In the bulk, the chemical decays with a rate $\gamma$, and is replenished with a background production rate $s K$, namely 
\begin{equation}
\frac{\dif c_\O}{\dif t}=s K+\Gamma \left(c_\A-c_\O\right)-\gamma c_\O,  
\end{equation}
which is equivalent to buffering the bulk concentration at a distant boundary. 

In stationary state (ss), the fixed-point structure of the dynamical system is determined by 
\begin{align}
    f(\sst{c}_\A) + s K &= 0,
\end{align}
which can be characterized via the discriminant $\Delta = 18 b^2 s (a+s)- 4 s (a+s)^3 + b^2 (a+s)^2 - 4 b^4 - 27 b^2 s^2$. Here, we have set $\gamma=0$ for simplicity of the presentation. For $\Delta <0$, the system has one fixed point, whereas for $\Delta >0$ it has three fixed points. Figure~\ref{fig:chem_dyn}\p{a} presents the corresponding bifurcation diagram showing how the fixed-points change as $s$ is varied. The phase-space flows of $c_\A$ and $c_\O$ at representative values of $s$ (Fig.~\ref{fig:chem_dyn}\p{a}) highlight how the fixed points change. Here, we demonstrate how the dynamics changes for different values of $s$ because this is perhaps the most accessible tuning parameter across a range of physical systems. The same level of control can be achieved by changing other parameters.

For $s=0$, the origin $c_\A=c_\O=0$ is always a stable fixed-point. Since the noise strength of the chemical dynamics is proportional to the square root of the chemical concentration (see Methods), the dynamics will be absorbed by a fixed-point at the origin, with no possibility for a noise-induced transition, as demonstrated in Fig.~\ref{fig:chem_dyn}\p{b}. Increasing $s$ moves the fixed-point away from the origin, preventing the dynamics from becoming fully absorbed (Fig.~\ref{fig:chem_dyn}\p{c}-\p{d}).
%

For finite values of $\gamma$, it is possible to create conditions such that the colloid can dynamically switch between consumer and producer states. The requirement on a non-zero $\gamma$ can be understood as follows. For the switch to be possible, the condition $\sst{c}_\O > \sst{c}_\A$ needs to hold at one stable fixed-point, whereas $\sst{c}_\O < \sst{c}_\A$ is required at the other stable fixed-point. However, with $\gamma=0$ we have $\sst{c}_\O = sK/\Gamma + \sst{c}_\A$, which does not allow these conditions to be met. This is the case in Fig.~\ref{fig:chem_dyn}\p{c}: although the system switches between fixed points, the average difference between $c_\O$ and $c_\A$ stays the same at each fixed point. On the other hand, in Fig.~\ref{fig:chem_dyn}\p{d} $c_\A$ can be seen to switch between being above or below $c_\O$ depending on the selected fixed-point. Thus, we require $s\neq0$ and $\gamma>0$ to access the full suite of behaviours.
The full analytical results for one colloid and the background with $\gamma\geq0$ are presented in the SI.

\begin{figure*}[t]
\centering
\includegraphics[width=\linewidth]{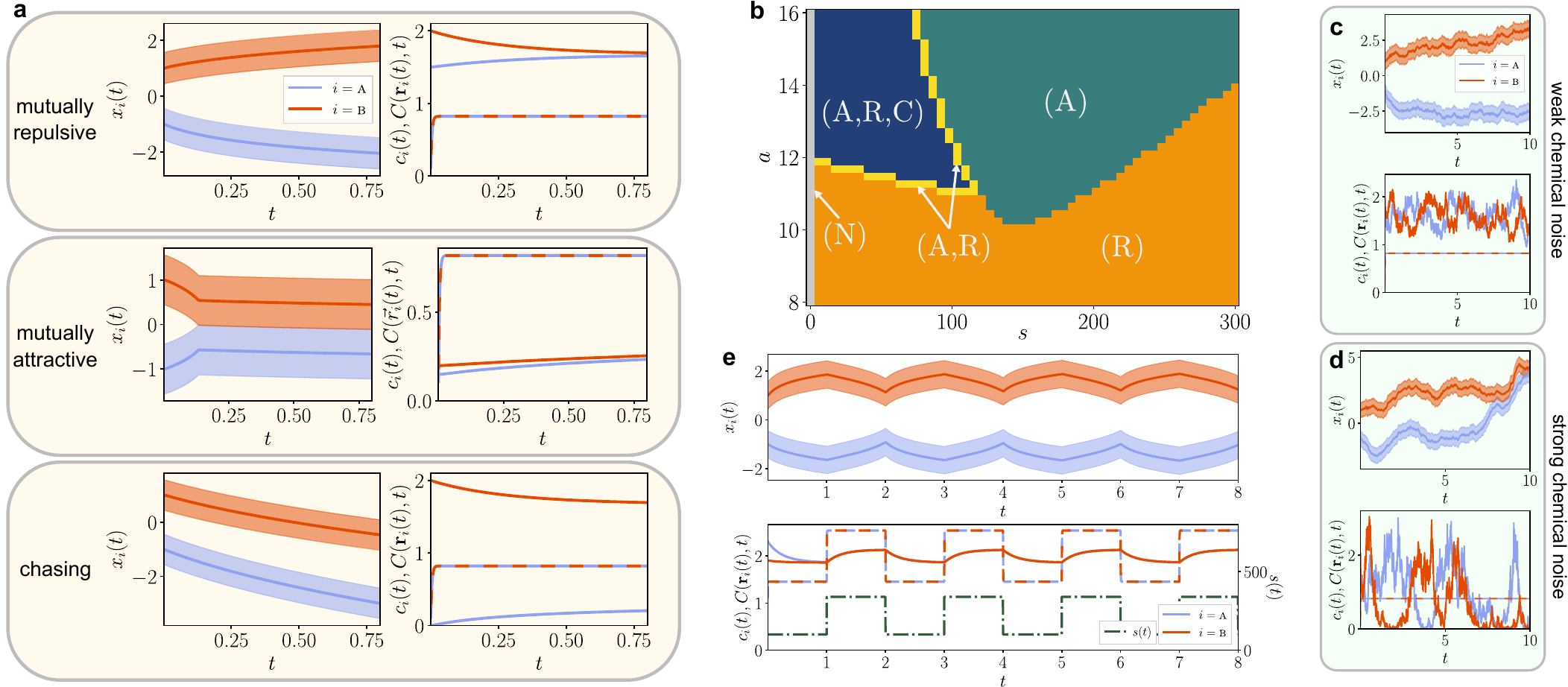}
\caption{\textbf{Interactions between two colloids.} 
\pc{a} Three distinct interaction combinations (mutually repulsive, mutually attractive, chasing) can be obtained from the same system by accessing different fixed points of the chemical dynamics. These trajectories show systems with the same parameters initiated at different internal chemical concentrations. $x_i(t)$ denotes colloid $i$'s centre of mass trajectory, with the shaded regions indicating the colloids' diameters. The internal chemical concentrations $c_i(t)$ are shown by solid lines, the value of the external chemical concentration at the colloids' positions $C(\vec{r}_i(t),t)$ are shown by dashed lines.
\pc{b} Classification of the interactions between two colloids at different values of $s$ and $a$. 
A: mutually attractive, R: mutually repulsive, C: chasing, N: no interactions.
\pc{c} Weak chemical noise does not significantly affect the interactions between the colloids. 
\pc{d} Strong chemical noise can trigger the colloids to spontaneously switch between being effective producers or consumers. This dynamically changes the nature of the interactions.
\pc{e} The colloids' interactions can be externally controlled. The dot-dashed line in the lower panel shows $s(t)$, which is periodically varied between a high and a low value. The colloids switch between mutual attraction and repulsion, depending on $s(t)$.
}
\label{fig:two_coll}
\end{figure*}

\begin{figure*}[t]
\centering
\includegraphics[width=\linewidth]{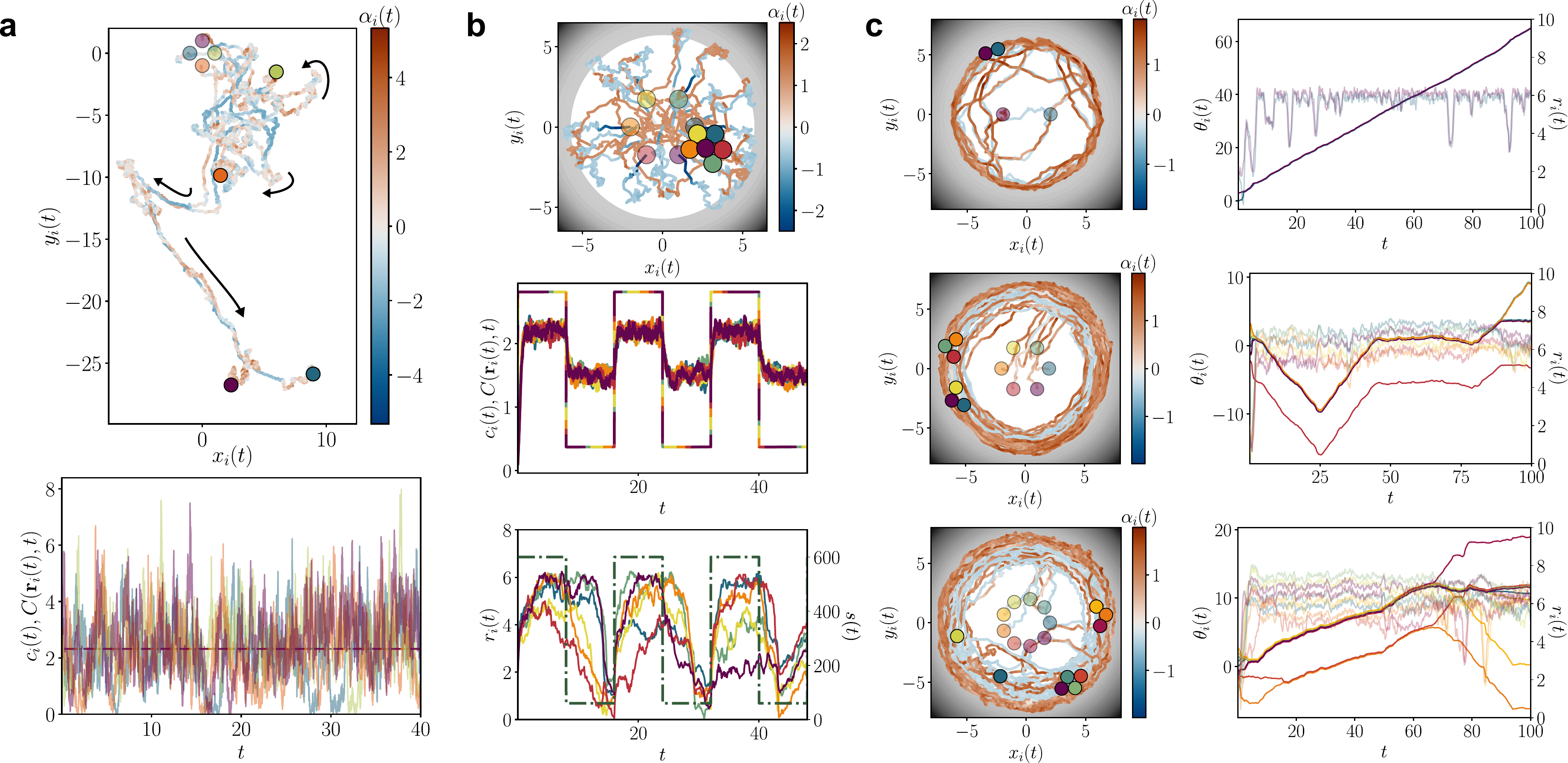}
\caption{\textbf{Collective behaviour of multiple interacting colloids.} 
\pc{a} Non-reciprocal interactions spontaneously emerge in a mixture of identically prepared colloids. Coloured circles mark the colloids' initial (light) and final (dark) positions. Note that although we use different colours to distinguish the colloids, each colloid is identically constructed. Coloured trails mark the colloids' trajectories, where the colour corresponds to the value of the effective chemical production rate $\alpha_i(t)=[c_i(t)-C(\vec{r}_i(t),t)]$ at that point of the trajectory. $c_i(t)$ and $C(\vec{r}_i(t),t)$ are plotted individually in the lower panel.
\pc{b} Varying $s$ externally can be used to control the interactions between the colloids. Here the colloids oscillate between attractive and repulsive interactions, which can be characterized by their radial coordinate plotted in the bottom panel, overlaying the $s(t)$ profile.
\pc{c} Colloids confined to a circular region exhibit persistent edge currents around the boundary. In the panels on the right we plot the trajectories in polar coordinates $(r_i(t),\theta_i(t))$.
}
\label{fig:many_coll}
\end{figure*}

The bistable chemical dynamics of two colloids with concentrations $c_\A$ and $c_\B$ exhibits interesting correlations that are mediated by the background medium concentration $c_\O$, giving rise to a range of possible chemical configurations. In stationary state, the equations for the fixed-points are given as
\begin{align}
   & f(\sst{c}_\A) + f(\sst{c}_\B) + s K = 0, \\
   & f(\sst{c}_\A) - f(\sst{c}_\B) = \Gamma \left(\sst{c}_\A - \sst{c}_\B\right),
\end{align}
for $\gamma=0$. 
In Fig.~\ref{fig:chem_dyn}\p{e}, we show how the fixed points of the chemical dynamics change as the external bulk chemical source strength $s$ is varied. Depending on the parameter values, the system can have one, two, or three unique stable fixed-points. A more detailed discussion of the series of bifurcations in the chemical dynamics is provided in the SI (including also for $\gamma>0$). 

A remarkable feature of the chemical dynamics with two colloids and the background is the possibility to have both symmetric ($c_\A=c_\B$) and asymmetric ($c_\A\neq c_\B$) stable fixed-points (see SI for the corresponding linear stability analysis). Let us consider an asymmetric fixed-point where $\sst{c}_\A<\sst{c}_\B$ without loss of generality. For the fixed-point to lead to chasing interactions, we require $\sst{c}_\A < \sst{c}_\O < \sst{c}_\B$, such that colloid $\A$ acquires a consumer state whilst colloid $\B$ becomes a producer. This condition is satisfied when
\begin{equation}
   \gamma\sst{c}_\A - \Gamma(\sst{c}_\B-\sst{c}_\A) < s K < \gamma\sst{c}_\B + \Gamma(\sst{c}_\B  - \sst{c}_\A),
\end{equation}
holds. We observe that such asymmetric fixed-points can be realized for an intermediate range of $s$, leading to the emergence of chasing interactions. 



\subsection{Cooperativity and tuning}


To study the mechanical interaction between colloids we need to incorporate the spatial dynamics of the background chemical, which diffuses in the medium with diffusion coefficient $\Dc$, and the phoretic response of the colloids to gradients in the bulk chemical concentration with mobility $\mu$ (Fig.~\ref{fig:cartoon}\p{b}), even though the space-free chemical dynamics provides an important indicator for the behaviours accessible in the full system with space included. We can classify the interactions between two colloids as mutually attractive, mutually repulsive, and chasing (Fig.~\ref{fig:cartoon}\p{c}). For chasing interactions, particle $\A$ is attracted by $\B$ whilst $\B$ is repelled by $\A$ (or vice versa). 
These three interaction types are demonstrated through particle trajectories in Fig.~\ref{fig:two_coll}\p{a}, which are obtained by solving the governing dynamical equations for colloids with diameter $\sigma$ that experience the phoretic effects arising from chemical activity and exert short-range repulsive interactions (see Methods). The trajectories shown in the figure are all obtained for the same system, but with different initial chemical concentrations. This highlights the flexibility of the proposed system, in which the same colloids can exhibit completely different interactions, simply by having the chemical variables at a different fixed-point of the chemical dynamics. 
To study the basic set of interactions, we set the positional noise to zero ($\Dp=0$), thereby restricting the positions of the colloids to one dimension, namely, $\vec{r}_i(t) = (x_i(t),0)$.

The versatility of the system can be observed from the range of interaction scenarios that can be realized by varying a minimal set of tuning parameters. In Fig.~\ref{fig:two_coll}\p{b}, we classify the range of interactions possible between two colloids for systems with different internal production rates $a$ and external source strengths $s$. We identify regimes with no interactions, purely attractive interactions, purely repulsive interactions, as well as a regime where both attractive and repulsive interactions are possible, and a regime where the interactions can be attractive, repulsive, or chasing. Interestingly, when $a\gtrsim 11$, varying $s$ alone explores the complete set of possible interaction combinations. The consequences of such a wide array of interaction types being accessible just by tuning one parameter are explored further when we discuss external control. 
These interaction combinations correspond to those predicted by the fixed-point analysis of the space-free chemical dynamics shown in Fig.~\ref{fig:chem_dyn}\p{e}. The cases where more than one interaction type is possible correspond to situations where the chemical dynamics has multiple fixed points. 
It is remarkable that the qualitative behavioural transitions as predicted by the space-free chemical dynamics are all observed in the full spatially resolved analysis of the system, as demonstrated by Fig.~\ref{fig:chem_dyn}\p{e} and Fig.~\ref{fig:two_coll}\p{a}-\p{b}.

It is important to examine the effect of introducing noise into the spatial and chemical dynamics. As shown in Fig.~\ref{fig:two_coll}\p{c}, weak chemical noise leaves the colloids with the same effective interactions. Strong chemical noise, however, can induce spontaneous transitions in the types of interactions between the colloids, switching from mutually attractive to chasing, or switching the chasing direction for example. This can be observed in Fig.~\ref{fig:two_coll}\p{d}.
Introducing positional noise does not significantly influence the chemical interactions, as can be seen in Fig.~\ref{fig:two_coll}\p{c}-\p{d} and in Fig.~\ref{fig:many_coll}, where we examine the interactions between multiple colloids.


We now demonstrate how the chemical dynamics can be influenced externally, in order to exploit the flexibility of the colloids' interactions.
In principle many of the system parameters could be controlled externally. We opt to vary the external bulk chemical source strength $s$ as this should be readily available to control for most systems, for example by buffering the concentration at the system boundaries.
To deterministically change the nature of the colloids' interactions, a bifurcation in the chemical dynamics can be triggered to force the chemical dynamics to a different fixed-point. This provides a robust change to the dynamics. A bifurcation diagram is shown for the space-free chemical dynamics in Fig.~\ref{fig:chem_dyn}\p{e}, and the different resulting interaction combinations for the full spatial system is plotted in Fig.~\ref{fig:two_coll}\p{b}. 
From these diagrams, we find that varying $s$ can lead to a change in the fixed-points.
For example, in Fig.~\ref{fig:chem_dyn}\p{e} we observe that for sufficiently large $s$ the system has one stable fixed-point, which can be tuned to transition from $\sst{c}_\A,\sst{c}_\B > \sst{c}_\O$ to $\sst{c}_\A,\sst{c}_\B < \sst{c}_\O$ as $s$ is increased. In this case, the interaction motif of the colloids switches from mutual attraction to mutual repulsion (or vice versa for $\mu<0$). This behaviour is confirmed in the full spatially resolved dynamics of the system, as seen in Fig.~\ref{fig:two_coll}\p{b}. The impact of this tunability is demonstrated in Fig.~\ref{fig:two_coll}\p{e}, where the behaviour of the system is controlled by periodically varying $s(t)$ between high and low values. 
The external chemical field responds quickly to the value of $s(t)$, jumping above or below the colloids' internal chemical concentrations and causing the colloids to switch between mutual attraction and mutual repulsion. 
One could also exploit the hysteresis observed in Fig.~\ref{fig:chem_dyn}, temporarily changing $s$ to push the chemical dynamics to a different fixed-point, before returning $s$ back to its previous value. With weak chemical noise, the chemical dynamics will remain at the target fixed-point. The ability to robustly and dynamically control the sign of the interactions between colloids is a remarkable feature of this system that will be useful in applications.

\subsection{Collective behaviour of many particles}

The interplay between the dynamics of the internal chemical state and the positions of colloidal particles has immense potential for the emergence of novel collective properties in a suspension of such particles, such as comet-like swarming that originates from spontaneously adopted chasing interactions that has been shown to emerge in thermophoretically active colloids \cite{Cohen2014}. In Fig.~\ref{fig:many_coll}, we highlight examples of such collective modes from a system with multiple colloids. We initialize the system with the colloids in identical states, and follow how the chemical dynamics can spontaneously trigger a range of interactions to emerge between the colloids. This can lead to emergent non-reciprocal chasing interactions between the colloids, as shown in Fig.~\ref{fig:many_coll}\p{a} and Supplementary Movie 1. We also observe the colloids pulsating towards and away from each other, as the colloids spontaneously switch between attractive and repulsive interactions.

In Fig.~\ref{fig:many_coll}\p{b} and Supplementary Movie 2, we demonstrate how the interactions between many colloids can be controlled by externally influencing the chemical dynamics. The bulk chemical source strength $s$ is periodically varied, which can be used to `quench' the interactions, for example, triggering all the colloids to attract each other. We observe the colloids switching between collective attraction and collective repulsion depending on the value of $s(t)$.

Chemically interacting colloids can form small clusters of `colloidal molecules', which can exhibit self-propulsion due to non-reciprocal interactions \cite{Soto2014PRL,Soto2015PRE,Agudo-Canalejo2019PRL,vanKesteren2023PNAS,Boniface2024NC,Peng2024SM}. In Fig.~\ref{fig:many_coll}\p{c} and Supplementary Movies 3-5, we show how multiple colloids confined by a circular wall interact. The system is initialized with identical colloids, which spontaneously form self-propelled clusters giving rise to emergent polarity in an intrinsically apolar system. When the self-propelled clusters encounter the confining wall, they spontaneously select one of the two directions to move along the boundary, leading to long-lived spontaneous symmetry-breaking chiral currents, as commonly observed in active matter \cite{Wioland2013,Wu2017}. The clusters break apart and recombine dynamically at relatively long times, based on the chemical and spatial dynamics as driven by noise. Increasing the number of colloids in the system decreases the lifetime of the self-propelled clusters, since with more colloids there are more interaction events to disrupt the spontaneous polarity of a cluster. These features can be observed in Fig.~\ref{fig:many_coll}\p{c}, where the trajectories of the colloids are shown in polar coordinates, with the radial coordinates $r_i(t)$ plotted by the light curves behind the winding angle around the origin $\theta_i(t)$ (see SI). In general the colloids accumulate at the confining boundary, with $r_i(t)\approx \Rbdr$ for most of the time. The chiral currents can be seen by the linear runs in the winding angle.

\section{Discussion}

We have shown how a mixture of identical chemically active colloids is able to spontaneously exhibit non-reciprocal chasing interactions due to bistable internal chemical dynamics that allows the particles to reversibly switch between acting as producers or consumers of a chemical. This novel mechanism for non-reciprocity in a mixture of identical particles occurs with completely shape-symmetric particles. Instead, the symmetry is broken in the chemical dynamics.
Each colloid acts as a small reaction chamber, with a semi-permeable membrane allowing for chemical to be exchanged between the colloids and environment, whilst separating the bistable chemical dynamics within each colloid from the surroundings. This allows for strong gradients in the chemical concentration seen by the catalysts inside two different colloids, allowing for stable and robust non-reciprocal interactions. If the chemical production were simply on the colloids surface, two nearby colloids would experience a very similar bulk chemical concentration due to the comparably fast diffusion of the chemical.

In order to highlight the novel chemical mechanisms, we have focussed here on a single species of colloid interacting via a single chemical species. Our work can be readily extended to include multiple species with the ability to control them individually via their specific reaction pathways. Therefore, we envisage that our methodology can be used to design cell-like reaction chambers with internal control, regulation, sensing, and the ability to perform mechanical functions.

In view of the advances in synthetic active colloids and vesicles, the mechanisms presented in this work should be readily realizable in experiments. As compared to many currently used protocols for single-species non-reciprocity such as vision cones, which heavily rely on external electronic sensing and information-processing machinery, the chemical symmetry-breaking mechanism we rely on here can be more naturally realized in microscopic colloidal systems and implemented with full autonomy. This capability has important implications for chemical and biomedical engineering applications such as drug delivery \cite{Chen2025}. 

The versatility of the colloids, each individually able to switch between acting as an effective producer or consumer of chemical, has powerful potential applications. An identical batch of colloids can be produced, that can be employed to perform completely opposite functions as desired. We also demonstrate how the colloids can be manipulated in situ, for example by temporarily flushing the system with chemical to drive each colloid to a particular state. By triggering bifurcations in the chemical dynamics, these desired behavioural changes can be performed very robustly.

\section{Methods}

\subsection{Equations of motion}
We consider a system of $N$ particles in a $d$-dimensional system. The equation of motion for particle $i$ is derived from an over-damped Langevin equation
\begin{align}
\begin{split}
    \frac{\dif{\vec{r}_i}}{\dif{t}} =  &-\mu \vec{\nabla} C(\vec{r}_i, t) - \frac{1}{\zeta}\vec{\nabla}_{\vec{r}_i}\sum_{j\neq i}U_\text{WCA}(r_{ij}) \\&- \frac{1}{\zeta}\vec{\nabla}_{\vec{r}_i}U_\text{C}(r_i) + \sqrt{2\Dp}\,\vec{\xi}_i(t), 
\end{split}\label{eq:eom_r}
\end{align}
where $\vec{\xi}_i(t)$ represents a Gaussian white noise with zero mean and unit strength. Particle volume exclusion is implemented using $U_\text{WCA}$, a Weeks-Chandler-Andersen (WCA) potential \cite{Weeks1971JCP} of the form
\begin{equation}
    U_\text{WCA}(r_{ij}) = \begin{cases}
        4\epsilon\left[\left(\frac{\sigma}{r_{ij}}\right)^{12} - \left(\frac{\sigma}{r_{ij}}\right)^{6}\right]+\epsilon,& r_{ij}\leq 2^{1/6} \sigma, \\
        0,& \text{otherwise}.
    \end{cases}
\end{equation}
The separation between particles $i$ and $j$ is denoted $r_{ij}=|\vec{r}_j-\vec{r}_i|$. For parts of the calculations, we also include a confinement potential $U_C$, which represents a circular confinement with radius $\Rbdr$, of the following truncated harmonic form
\begin{equation}
    U_\text{H}(r_i) = \begin{cases}
        0,& r_{i} < \Rbdr,\\
        \frac{1}{2}k(r_i-\Rbdr)^2,& r_{i} \geq \Rbdr.
    \end{cases}
\end{equation}

The dynamics of the chemical field $C$ obeys the following stochastic partial differential equation
\begin{align}
\begin{split}
 \hskip-2mm   \partial_t C(\vec{x},t) &= \Dc \nabla^2 C(\vec{x},t) - \gamma C(\vec{x},t) + s(t) K \\&+ \Gamma\sum_{i} \big[c_i(t)-C(\vec{r}_i(t),t)\big] V_d \delta^{d}(\vec{x}-\vec{r}_i(t)), 
\end{split}\label{eq:eom_C}
\end{align}
where $\Dc$ is the diffusion coefficient of the chemical in the bulk, and the chemical degrades with a decay rate $\gamma$. $s(t)$ is an externally imposed spatially uniform source, for example arising from buffering the bulk concentration at a distant boundary \cite{Golestanian2015PRL}.

Each colloid is treated as a reaction chamber, with colloid $i$ containing a chemical concentration $c_i$. The chemical is exchanged between each colloid's internal reservoir and the bulk with rate $\Gamma$. The internal chemical dynamics evolves as
\begin{equation}
    \partial_t c_i(t) = f(c_i(t)) + \Gamma\big[C(\vec{r}_i(t),t)-c_i(t)\big] + \sqrt{2Q c_i(t)} \,\xi_i(t), \label{eq:eom_c}
\end{equation}
where $\xi_i(t)$ represents Gaussian white noise with zero mean and unit strength, and $Q$ is the strength of the chemical noise that is implemented in the multiplicative form that is characteristic of Poissonian number fluctuations. 
For $f(c)$, we will employ a Hill function, combined with a decay term, namely
\begin{equation}
    f(c) = \frac{a  K \,c^2}{K^2+c^2} - b c. \label{eq:f}
\end{equation}
While we use the Hill coefficient of 2 to represent a basic form of cooperative reaction that can arise from a reaction network, the phenomenology is robust with respect to using other values for the Hill coefficient, or, indeed, any bistable rate function $f(c)$.

\subsection{Numerical integration}
To integrate the stochastic equations of motion, we start by deriving the formal solution for the bulk chemical field. This is obtained by combining the actual sources with the Green function solution to Eq.~\eqref{eq:eom_C} with a point source, resulting in
\begin{align}
\begin{split}
C(\vec{x},t) &= C_0 + K \int_0^t \dif{t'}\, e^{-\gamma (t-t')}s(t') \\ &+ \sum_i \int^t_0  \dif{t'}\Bigg\{ \frac{\Gamma \, e^{-\gamma (t-t')}}{\left[4\pi\Dc (t-t')\right]^{d/2}}\\ & \times \exp{\left(-\frac{|\vec{x}-\vec{r}_i(t')|^2}{4\Dc (t-t')}\right)} \, \big[c_i(t')-C(\vec{r}_i(t'),t')\big]  \Bigg\}, \label{eq:Greens_C}
\end{split}
\end{align}
where $C_0$ represents an initial uniform background concentration.

The solution for $C(\vec{x},t)$ is not closed since it depends on the particle trajectories $\vec{r}_i(t)$ and the internal chemical dynamics $c_i(t)$. We can, however, combine this formal solution and its spatial gradient with the ordinary stochastic differential equations given in Eq.~\eqref{eq:eom_r} and Eq.~\eqref{eq:eom_c}, and numerically integrate the resulting system of equations.
At each time-step, $C(\vec{r}_i(t),t)$ and $\vec{\nabla}C(\vec{r}_i(t),t)$ are calculated at every particle's position, using the particle positions at each previous time-step to numerically evaluate the integrals.
Given $C(\vec{r}_i(t),t)$ and $\vec{\nabla}C(\vec{r}_i(t),t)$, each particle's position $\vec{r}_i$ and internal chemical concentration $c_i$ can be updated using an Euler-Maruyama scheme, 
with a time-step $dt=\sigma^2/(2\Dc)=0.001$ that accounts for the finite size of the colloids in the Green function integral.
In the presence of decay ($\gamma>0$), the integrals can be truncated, integrating from $t-\frac{\cutoff}{\gamma}$ to $t$, where $\cutoff$ is a cut-off multiple for which $e^{-\cutoff}$ is deemed to be sufficiently small such that we can ignore all contributions to the integral for $t'<t-\frac{\cutoff}{\gamma}$. We use $\cutoff=5$.

\subsection{Space-free chemical dynamics}
To analyze the fixed-points of the chemical dynamics, we employ a mean-field model with no spatial dependence. The bulk chemical field is represented by a single space-independent variable $c_\O(t)$, where formally
\begin{equation}
    c_\O(t) = \frac{1}{V} \int_V \dif^d{\vec{x}} \, C(\vec{x},t).
\end{equation}
In the mean-field limit, 
this concentration is taken to be seen equally by all particles independently of their positions. This leads to the following dynamical equations for the chemical dynamics 
\begin{align}
    \partial_t c_\O &= s K + \Gamma \sum_{i} (c_i-c_\O) - \gamma c_\O, \\
    \partial_t c_i &= f(c_i) + \Gamma(c_\O-c_i) + \sqrt{2Q c_i(t)} \,\xi_i(t),
\end{align}
where $i$ labels $N$ different colloids.

In the SI, we present the fixed-point structure of these equations, and analyze their stability (analytically for $N=1$, numerically for $N>1$).
The space-free equations result in slightly different quantitive results for the fixed-points as compared with the full spatially resolved system, although qualitatively the same features are observed.

\bibliography{refs_JM,Golestanian,refs_RG}

\begin{thebibliography}{8}%
\makeatletter
\providecommand \@ifxundefined [1]{%
 \@ifx{#1\undefined}
}%
\providecommand \@ifnum [1]{%
 \ifnum #1\expandafter \@firstoftwo
 \else \expandafter \@secondoftwo
 \fi
}%
\providecommand \@ifx [1]{%
 \ifx #1\expandafter \@firstoftwo
 \else \expandafter \@secondoftwo
 \fi
}%
\providecommand \natexlab [1]{#1}%
\providecommand \enquote  [1]{``#1''}%
\providecommand \bibnamefont  [1]{#1}%
\providecommand \bibfnamefont [1]{#1}%
\providecommand \citenamefont [1]{#1}%
\providecommand \href@noop [0]{\@secondoftwo}%
\providecommand \href [0]{\begingroup \@sanitize@url \@href}%
\providecommand \@href[1]{\@@startlink{#1}\@@href}%
\providecommand \@@href[1]{\endgroup#1\@@endlink}%
\providecommand \@sanitize@url [0]{\catcode `\\12\catcode `\$12\catcode
  `\&12\catcode `\#12\catcode `\^12\catcode `\_12\catcode `\%12\relax}%
\providecommand \@@startlink[1]{}%
\providecommand \@@endlink[0]{}%
\providecommand \url  [0]{\begingroup\@sanitize@url \@url }%
\providecommand \@url [1]{\endgroup\@href {#1}{\urlprefix }}%
\providecommand \urlprefix  [0]{URL }%
\providecommand \Eprint [0]{\href }%
\providecommand \doibase [0]{https://doi.org/}%
\providecommand \selectlanguage [0]{\@gobble}%
\providecommand \bibinfo  [0]{\@secondoftwo}%
\providecommand \bibfield  [0]{\@secondoftwo}%
\providecommand \translation [1]{[#1]}%
\providecommand \BibitemOpen [0]{}%
\providecommand \bibitemStop [0]{}%
\providecommand \bibitemNoStop [0]{.\EOS\space}%
\providecommand \EOS [0]{\spacefactor3000\relax}%
\providecommand \BibitemShut  [1]{\csname bibitem#1\endcsname}%
\let\auto@bib@innerbib\@empty
\bibitem [{\citenamefont {Meurer}\ \emph {et~al.}(2017)\citenamefont {Meurer},
  \citenamefont {Smith}, \citenamefont {Paprocki}, \citenamefont {{\v
  C}ert{\'i}k}, \citenamefont {Kirpichev}, \citenamefont {Rocklin},
  \citenamefont {Kumar}, \citenamefont {Ivanov}, \citenamefont {Moore},
  \citenamefont {Singh}, \citenamefont {Rathnayake}, \citenamefont {Vig},
  \citenamefont {Granger}, \citenamefont {Muller}, \citenamefont {Bonazzi},
  \citenamefont {Gupta}, \citenamefont {Vats}, \citenamefont {Johansson},
  \citenamefont {Pedregosa}, \citenamefont {Curry}, \citenamefont {Terrel},
  \citenamefont {Rou{\v c}ka}, \citenamefont {Saboo}, \citenamefont {Fernando},
  \citenamefont {Kulal}, \citenamefont {Cimrman},\ and\ \citenamefont
  {Scopatz}}]{SymPy}%
  \BibitemOpen
  \bibfield  {author} {\bibinfo {author} {\bibfnamefont {A.}~\bibnamefont
  {Meurer}}, \bibinfo {author} {\bibfnamefont {C.~P.}\ \bibnamefont {Smith}},
  \bibinfo {author} {\bibfnamefont {M.}~\bibnamefont {Paprocki}}, \bibinfo
  {author} {\bibfnamefont {O.}~\bibnamefont {{\v C}ert{\'i}k}}, \bibinfo
  {author} {\bibfnamefont {S.~B.}\ \bibnamefont {Kirpichev}}, \bibinfo {author}
  {\bibfnamefont {M.}~\bibnamefont {Rocklin}}, \bibinfo {author} {\bibfnamefont
  {{\relax Am}.}~\bibnamefont {Kumar}}, \bibinfo {author} {\bibfnamefont
  {S.}~\bibnamefont {Ivanov}}, \bibinfo {author} {\bibfnamefont {J.~K.}\
  \bibnamefont {Moore}}, \bibinfo {author} {\bibfnamefont {S.}~\bibnamefont
  {Singh}}, \bibinfo {author} {\bibfnamefont {T.}~\bibnamefont {Rathnayake}},
  \bibinfo {author} {\bibfnamefont {S.}~\bibnamefont {Vig}}, \bibinfo {author}
  {\bibfnamefont {B.~E.}\ \bibnamefont {Granger}}, \bibinfo {author}
  {\bibfnamefont {R.~P.}\ \bibnamefont {Muller}}, \bibinfo {author}
  {\bibfnamefont {F.}~\bibnamefont {Bonazzi}}, \bibinfo {author} {\bibfnamefont
  {H.}~\bibnamefont {Gupta}}, \bibinfo {author} {\bibfnamefont
  {S.}~\bibnamefont {Vats}}, \bibinfo {author} {\bibfnamefont {F.}~\bibnamefont
  {Johansson}}, \bibinfo {author} {\bibfnamefont {F.}~\bibnamefont
  {Pedregosa}}, \bibinfo {author} {\bibfnamefont {M.~J.}\ \bibnamefont
  {Curry}}, \bibinfo {author} {\bibfnamefont {A.~R.}\ \bibnamefont {Terrel}},
  \bibinfo {author} {\bibfnamefont {{\v S}.}~\bibnamefont {Rou{\v c}ka}},
  \bibinfo {author} {\bibfnamefont {A.}~\bibnamefont {Saboo}}, \bibinfo
  {author} {\bibfnamefont {I.}~\bibnamefont {Fernando}}, \bibinfo {author}
  {\bibfnamefont {S.}~\bibnamefont {Kulal}}, \bibinfo {author} {\bibfnamefont
  {R.}~\bibnamefont {Cimrman}},\ and\ \bibinfo {author} {\bibfnamefont
  {A.}~\bibnamefont {Scopatz}},\ }\bibfield  {title} {\bibinfo {title}
  {{{SymPy}}: Symbolic computing in {{Python}}},\ }\href
  {https://doi.org/10.7717/peerj-cs.103} {\bibfield  {journal} {\bibinfo
  {journal} {PeerJ Computer Science}\ }\textbf {\bibinfo {volume} {3}},\
  \bibinfo {pages} {e103} (\bibinfo {year} {2017})}\BibitemShut {NoStop}%
\bibitem [{\citenamefont {Harris}\ \emph {et~al.}(2020)\citenamefont {Harris},
  \citenamefont {Millman}, \citenamefont {{van der Walt}}, \citenamefont
  {Gommers}, \citenamefont {Virtanen}, \citenamefont {Cournapeau},
  \citenamefont {Wieser}, \citenamefont {Taylor}, \citenamefont {Berg},
  \citenamefont {Smith}, \citenamefont {Kern}, \citenamefont {Picus},
  \citenamefont {Hoyer}, \citenamefont {{van Kerkwijk}}, \citenamefont {Brett},
  \citenamefont {Haldane}, \citenamefont {{del R{\'i}o}}, \citenamefont
  {Wiebe}, \citenamefont {Peterson}, \citenamefont {{G{\'e}rard-Marchant}},
  \citenamefont {Sheppard}, \citenamefont {Reddy}, \citenamefont {Weckesser},
  \citenamefont {Abbasi}, \citenamefont {Gohlke},\ and\ \citenamefont
  {Oliphant}}]{NumPy}%
  \BibitemOpen
  \bibfield  {author} {\bibinfo {author} {\bibfnamefont {C.~R.}\ \bibnamefont
  {Harris}}, \bibinfo {author} {\bibfnamefont {K.~J.}\ \bibnamefont {Millman}},
  \bibinfo {author} {\bibfnamefont {S.~J.}\ \bibnamefont {{van der Walt}}},
  \bibinfo {author} {\bibfnamefont {R.}~\bibnamefont {Gommers}}, \bibinfo
  {author} {\bibfnamefont {P.}~\bibnamefont {Virtanen}}, \bibinfo {author}
  {\bibfnamefont {D.}~\bibnamefont {Cournapeau}}, \bibinfo {author}
  {\bibfnamefont {E.}~\bibnamefont {Wieser}}, \bibinfo {author} {\bibfnamefont
  {J.}~\bibnamefont {Taylor}}, \bibinfo {author} {\bibfnamefont
  {S.}~\bibnamefont {Berg}}, \bibinfo {author} {\bibfnamefont {N.~J.}\
  \bibnamefont {Smith}}, \bibinfo {author} {\bibfnamefont {R.}~\bibnamefont
  {Kern}}, \bibinfo {author} {\bibfnamefont {M.}~\bibnamefont {Picus}},
  \bibinfo {author} {\bibfnamefont {S.}~\bibnamefont {Hoyer}}, \bibinfo
  {author} {\bibfnamefont {M.~H.}\ \bibnamefont {{van Kerkwijk}}}, \bibinfo
  {author} {\bibfnamefont {M.}~\bibnamefont {Brett}}, \bibinfo {author}
  {\bibfnamefont {A.}~\bibnamefont {Haldane}}, \bibinfo {author} {\bibfnamefont
  {J.~F.}\ \bibnamefont {{del R{\'i}o}}}, \bibinfo {author} {\bibfnamefont
  {M.}~\bibnamefont {Wiebe}}, \bibinfo {author} {\bibfnamefont
  {P.}~\bibnamefont {Peterson}}, \bibinfo {author} {\bibfnamefont
  {P.}~\bibnamefont {{G{\'e}rard-Marchant}}}, \bibinfo {author} {\bibfnamefont
  {K.}~\bibnamefont {Sheppard}}, \bibinfo {author} {\bibfnamefont
  {T.}~\bibnamefont {Reddy}}, \bibinfo {author} {\bibfnamefont
  {W.}~\bibnamefont {Weckesser}}, \bibinfo {author} {\bibfnamefont
  {H.}~\bibnamefont {Abbasi}}, \bibinfo {author} {\bibfnamefont
  {C.}~\bibnamefont {Gohlke}},\ and\ \bibinfo {author} {\bibfnamefont {T.~E.}\
  \bibnamefont {Oliphant}},\ }\bibfield  {title} {\bibinfo {title} {Array
  programming with {{NumPy}}},\ }\href
  {https://doi.org/10.1038/s41586-020-2649-2} {\bibfield  {journal} {\bibinfo
  {journal} {Nature}\ }\textbf {\bibinfo {volume} {585}},\ \bibinfo {pages}
  {357} (\bibinfo {year} {2020})}\BibitemShut {NoStop}%
\bibitem [{\citenamefont {Virtanen}\ \emph {et~al.}(2020)\citenamefont
  {Virtanen}, \citenamefont {Gommers}, \citenamefont {Oliphant}, \citenamefont
  {Haberland}, \citenamefont {Reddy}, \citenamefont {Cournapeau}, \citenamefont
  {Burovski}, \citenamefont {Peterson}, \citenamefont {Weckesser},
  \citenamefont {Bright}, \citenamefont {{van der Walt}}, \citenamefont
  {Brett}, \citenamefont {Wilson}, \citenamefont {Millman}, \citenamefont
  {Mayorov}, \citenamefont {Nelson}, \citenamefont {Jones}, \citenamefont
  {Kern}, \citenamefont {Larson}, \citenamefont {Carey}, \citenamefont {Polat},
  \citenamefont {Feng}, \citenamefont {Moore}, \citenamefont {VanderPlas},
  \citenamefont {Laxalde}, \citenamefont {Perktold}, \citenamefont {Cimrman},
  \citenamefont {Henriksen}, \citenamefont {Quintero}, \citenamefont {Harris},
  \citenamefont {Archibald}, \citenamefont {Ribeiro}, \citenamefont
  {Pedregosa}, \citenamefont {{van Mulbregt}},\ and\ \citenamefont {{SciPy 1.0
  Contributors}}}]{SciPy}%
  \BibitemOpen
  \bibfield  {author} {\bibinfo {author} {\bibfnamefont {P.}~\bibnamefont
  {Virtanen}}, \bibinfo {author} {\bibfnamefont {R.}~\bibnamefont {Gommers}},
  \bibinfo {author} {\bibfnamefont {T.~E.}\ \bibnamefont {Oliphant}}, \bibinfo
  {author} {\bibfnamefont {M.}~\bibnamefont {Haberland}}, \bibinfo {author}
  {\bibfnamefont {T.}~\bibnamefont {Reddy}}, \bibinfo {author} {\bibfnamefont
  {D.}~\bibnamefont {Cournapeau}}, \bibinfo {author} {\bibfnamefont
  {E.}~\bibnamefont {Burovski}}, \bibinfo {author} {\bibfnamefont
  {P.}~\bibnamefont {Peterson}}, \bibinfo {author} {\bibfnamefont
  {W.}~\bibnamefont {Weckesser}}, \bibinfo {author} {\bibfnamefont
  {J.}~\bibnamefont {Bright}}, \bibinfo {author} {\bibfnamefont {S.~J.}\
  \bibnamefont {{van der Walt}}}, \bibinfo {author} {\bibfnamefont
  {M.}~\bibnamefont {Brett}}, \bibinfo {author} {\bibfnamefont
  {J.}~\bibnamefont {Wilson}}, \bibinfo {author} {\bibfnamefont {K.~J.}\
  \bibnamefont {Millman}}, \bibinfo {author} {\bibfnamefont {N.}~\bibnamefont
  {Mayorov}}, \bibinfo {author} {\bibfnamefont {A.~R.~J.}\ \bibnamefont
  {Nelson}}, \bibinfo {author} {\bibfnamefont {E.}~\bibnamefont {Jones}},
  \bibinfo {author} {\bibfnamefont {R.}~\bibnamefont {Kern}}, \bibinfo {author}
  {\bibfnamefont {E.}~\bibnamefont {Larson}}, \bibinfo {author} {\bibfnamefont
  {C.~J.}\ \bibnamefont {Carey}}, \bibinfo {author} {\bibfnamefont
  {{\.I}.}~\bibnamefont {Polat}}, \bibinfo {author} {\bibfnamefont
  {Y.}~\bibnamefont {Feng}}, \bibinfo {author} {\bibfnamefont {E.~W.}\
  \bibnamefont {Moore}}, \bibinfo {author} {\bibfnamefont {J.}~\bibnamefont
  {VanderPlas}}, \bibinfo {author} {\bibfnamefont {D.}~\bibnamefont {Laxalde}},
  \bibinfo {author} {\bibfnamefont {J.}~\bibnamefont {Perktold}}, \bibinfo
  {author} {\bibfnamefont {R.}~\bibnamefont {Cimrman}}, \bibinfo {author}
  {\bibfnamefont {I.}~\bibnamefont {Henriksen}}, \bibinfo {author}
  {\bibfnamefont {E.~A.}\ \bibnamefont {Quintero}}, \bibinfo {author}
  {\bibfnamefont {C.~R.}\ \bibnamefont {Harris}}, \bibinfo {author}
  {\bibfnamefont {A.~M.}\ \bibnamefont {Archibald}}, \bibinfo {author}
  {\bibfnamefont {A.~H.}\ \bibnamefont {Ribeiro}}, \bibinfo {author}
  {\bibfnamefont {F.}~\bibnamefont {Pedregosa}}, \bibinfo {author}
  {\bibfnamefont {P.}~\bibnamefont {{van Mulbregt}}},\ and\ \bibinfo {author}
  {\bibnamefont {{SciPy 1.0 Contributors}}},\ }\bibfield  {title} {\bibinfo
  {title} {{{SciPy}} 1.0: {{Fundamental}} algorithms for scientific computing
  in python},\ }\href {https://doi.org/10.1038/s41592-019-0686-2} {\bibfield
  {journal} {\bibinfo  {journal} {Nature Methods}\ }\textbf {\bibinfo {volume}
  {17}},\ \bibinfo {pages} {261} (\bibinfo {year} {2020})}\BibitemShut
  {NoStop}%
\bibitem [{\citenamefont {Dormand}\ and\ \citenamefont
  {Prince}(1980)}]{Dormand1980JoCaAM}%
  \BibitemOpen
  \bibfield  {author} {\bibinfo {author} {\bibfnamefont {J.~R.}\ \bibnamefont
  {Dormand}}\ and\ \bibinfo {author} {\bibfnamefont {P.~J.}\ \bibnamefont
  {Prince}},\ }\bibfield  {title} {\bibinfo {title} {A family of embedded
  {{Runge-Kutta}} formulae},\ }\href
  {https://doi.org/10.1016/0771-050X(80)90013-3} {\bibfield  {journal}
  {\bibinfo  {journal} {Journal of Computational and Applied Mathematics}\
  }\textbf {\bibinfo {volume} {6}},\ \bibinfo {pages} {19} (\bibinfo {year}
  {1980})}\BibitemShut {NoStop}%
\bibitem [{\citenamefont {Wang}\ \emph {et~al.}(2023)\citenamefont {Wang},
  \citenamefont {Zhang}, \citenamefont {Chen}, \citenamefont {He},
  \citenamefont {Li},\ and\ \citenamefont {Wu}}]{BrainPy}%
  \BibitemOpen
  \bibfield  {author} {\bibinfo {author} {\bibfnamefont {C.}~\bibnamefont
  {Wang}}, \bibinfo {author} {\bibfnamefont {T.}~\bibnamefont {Zhang}},
  \bibinfo {author} {\bibfnamefont {X.}~\bibnamefont {Chen}}, \bibinfo {author}
  {\bibfnamefont {S.}~\bibnamefont {He}}, \bibinfo {author} {\bibfnamefont
  {S.}~\bibnamefont {Li}},\ and\ \bibinfo {author} {\bibfnamefont
  {S.}~\bibnamefont {Wu}},\ }\bibfield  {title} {\bibinfo {title} {{{BrainPy}},
  a flexible, integrative, efficient, and extensible framework for
  general-purpose brain dynamics programming},\ }\href
  {https://doi.org/10.7554/eLife.86365} {\bibfield  {journal} {\bibinfo
  {journal} {eLife}\ }\textbf {\bibinfo {volume} {12}},\ \bibinfo {pages}
  {e86365} (\bibinfo {year} {2023})}\BibitemShut {NoStop}%
\bibitem [{\citenamefont {Hunter}(2007)}]{Matplotlib}%
  \BibitemOpen
  \bibfield  {author} {\bibinfo {author} {\bibfnamefont {J.~D.}\ \bibnamefont
  {Hunter}},\ }\bibfield  {title} {\bibinfo {title} {Matplotlib: {{A 2D}}
  graphics environment},\ }\href {https://doi.org/10.1109/MCSE.2007.55}
  {\bibfield  {journal} {\bibinfo  {journal} {Computing in Science \&
  Engineering}\ }\textbf {\bibinfo {volume} {9}},\ \bibinfo {pages} {90}
  (\bibinfo {year} {2007})}\BibitemShut {NoStop}%
\bibitem [{\citenamefont {{van der Velden}}(2020)}]{CMasher}%
  \BibitemOpen
  \bibfield  {author} {\bibinfo {author} {\bibfnamefont {E.}~\bibnamefont {{van
  der Velden}}},\ }\bibfield  {title} {\bibinfo {title} {{{CMasher}}:
  {{Scientific}} colormaps for making accessible, informative and 'cmashing'
  plots},\ }\href {https://doi.org/10.21105/joss.02004} {\bibfield  {journal}
  {\bibinfo  {journal} {The Journal of Open Source Software}\ }\textbf
  {\bibinfo {volume} {5}},\ \bibinfo {pages} {2004} (\bibinfo {year} {2020})},\
  \Eprint {https://arxiv.org/abs/2003.01069} {arXiv:2003.01069 [eess.IV]}
  \BibitemShut {NoStop}%
\bibitem [{\citenamefont {Crameri}(2023)}]{Crameri2023}%
  \BibitemOpen
  \bibfield  {author} {\bibinfo {author} {\bibfnamefont {F.}~\bibnamefont
  {Crameri}},\ }\href {https://doi.org/10.5281/zenodo.8409685} {\bibinfo
  {title} {Scientific colour maps}},\ \bibinfo {howpublished} {Zenodo}
  (\bibinfo {year} {2023})\BibitemShut {NoStop}%
\end{thebibliography}%


\begin{thebibliography}{39}%
\makeatletter
\providecommand \@ifxundefined [1]{%
 \@ifx{#1\undefined}
}%
\providecommand \@ifnum [1]{%
 \ifnum #1\expandafter \@firstoftwo
 \else \expandafter \@secondoftwo
 \fi
}%
\providecommand \@ifx [1]{%
 \ifx #1\expandafter \@firstoftwo
 \else \expandafter \@secondoftwo
 \fi
}%
\providecommand \natexlab [1]{#1}%
\providecommand \enquote  [1]{``#1''}%
\providecommand \bibnamefont  [1]{#1}%
\providecommand \bibfnamefont [1]{#1}%
\providecommand \citenamefont [1]{#1}%
\providecommand \href@noop [0]{\@secondoftwo}%
\providecommand \href [0]{\begingroup \@sanitize@url \@href}%
\providecommand \@href[1]{\@@startlink{#1}\@@href}%
\providecommand \@@href[1]{\endgroup#1\@@endlink}%
\providecommand \@sanitize@url [0]{\catcode `\\12\catcode `\$12\catcode
  `\&12\catcode `\#12\catcode `\^12\catcode `\_12\catcode `\%12\relax}%
\providecommand \@@startlink[1]{}%
\providecommand \@@endlink[0]{}%
\providecommand \url  [0]{\begingroup\@sanitize@url \@url }%
\providecommand \@url [1]{\endgroup\@href {#1}{\urlprefix }}%
\providecommand \urlprefix  [0]{URL }%
\providecommand \Eprint [0]{\href }%
\providecommand \doibase [0]{https://doi.org/}%
\providecommand \selectlanguage [0]{\@gobble}%
\providecommand \bibinfo  [0]{\@secondoftwo}%
\providecommand \bibfield  [0]{\@secondoftwo}%
\providecommand \translation [1]{[#1]}%
\providecommand \BibitemOpen [0]{}%
\providecommand \bibitemStop [0]{}%
\providecommand \bibitemNoStop [0]{.\EOS\space}%
\providecommand \EOS [0]{\spacefactor3000\relax}%
\providecommand \BibitemShut  [1]{\csname bibitem#1\endcsname}%
\let\auto@bib@innerbib\@empty
\bibitem [{\citenamefont {Rousselet}\ \emph {et~al.}(1994)\citenamefont
  {Rousselet}, \citenamefont {Salome}, \citenamefont {Ajdari},\ and\
  \citenamefont {Prost}}]{Rousselet1994}%
  \BibitemOpen
  \bibfield  {author} {\bibinfo {author} {\bibfnamefont {J.}~\bibnamefont
  {Rousselet}}, \bibinfo {author} {\bibfnamefont {L.}~\bibnamefont {Salome}},
  \bibinfo {author} {\bibfnamefont {A.}~\bibnamefont {Ajdari}},\ and\ \bibinfo
  {author} {\bibfnamefont {J.}~\bibnamefont {Prost}},\ }\bibfield  {title}
  {\bibinfo {title} {Directional motion of brownian particles induced by a
  periodic asymmetric potential},\ }\href {https://doi.org/10.1038/370446a0}
  {\bibfield  {journal} {\bibinfo  {journal} {Nature}\ }\textbf {\bibinfo
  {volume} {370}},\ \bibinfo {pages} {446–447} (\bibinfo {year}
  {1994})}\BibitemShut {NoStop}%
\bibitem [{\citenamefont {Pumm}\ \emph {et~al.}(2022)\citenamefont {Pumm},
  \citenamefont {Engelen}, \citenamefont {Kopperger}, \citenamefont {Isensee},
  \citenamefont {Vogt}, \citenamefont {Kozina}, \citenamefont {Kube},
  \citenamefont {Honemann}, \citenamefont {Bertosin}, \citenamefont
  {Langecker}, \citenamefont {Golestanian}, \citenamefont {Simmel},\ and\
  \citenamefont {Dietz}}]{Pumm2022}%
  \BibitemOpen
  \bibfield  {author} {\bibinfo {author} {\bibfnamefont {A.-K.}\ \bibnamefont
  {Pumm}}, \bibinfo {author} {\bibfnamefont {W.}~\bibnamefont {Engelen}},
  \bibinfo {author} {\bibfnamefont {E.}~\bibnamefont {Kopperger}}, \bibinfo
  {author} {\bibfnamefont {J.}~\bibnamefont {Isensee}}, \bibinfo {author}
  {\bibfnamefont {M.}~\bibnamefont {Vogt}}, \bibinfo {author} {\bibfnamefont
  {V.}~\bibnamefont {Kozina}}, \bibinfo {author} {\bibfnamefont
  {M.}~\bibnamefont {Kube}}, \bibinfo {author} {\bibfnamefont {M.~N.}\
  \bibnamefont {Honemann}}, \bibinfo {author} {\bibfnamefont {E.}~\bibnamefont
  {Bertosin}}, \bibinfo {author} {\bibfnamefont {M.}~\bibnamefont {Langecker}},
  \bibinfo {author} {\bibfnamefont {R.}~\bibnamefont {Golestanian}}, \bibinfo
  {author} {\bibfnamefont {F.~C.}\ \bibnamefont {Simmel}},\ and\ \bibinfo
  {author} {\bibfnamefont {H.}~\bibnamefont {Dietz}},\ }\bibfield  {title}
  {\bibinfo {title} {A dna origami rotary ratchet motor},\ }\href
  {https://doi.org/10.1038/s41586-022-04910-y} {\bibfield  {journal} {\bibinfo
  {journal} {Nature}\ }\textbf {\bibinfo {volume} {607}},\ \bibinfo {pages}
  {492} (\bibinfo {year} {2022})}\BibitemShut {NoStop}%
\bibitem [{\citenamefont {Kruse}\ \emph {et~al.}(2004)\citenamefont {Kruse},
  \citenamefont {Joanny}, \citenamefont {J\"{u}licher}, \citenamefont {Prost},\
  and\ \citenamefont {Sekimoto}}]{Kruse2004}%
  \BibitemOpen
  \bibfield  {author} {\bibinfo {author} {\bibfnamefont {K.}~\bibnamefont
  {Kruse}}, \bibinfo {author} {\bibfnamefont {J.~F.}\ \bibnamefont {Joanny}},
  \bibinfo {author} {\bibfnamefont {F.}~\bibnamefont {J\"{u}licher}}, \bibinfo
  {author} {\bibfnamefont {J.}~\bibnamefont {Prost}},\ and\ \bibinfo {author}
  {\bibfnamefont {K.}~\bibnamefont {Sekimoto}},\ }\bibfield  {title} {\bibinfo
  {title} {Asters, vortices, and rotating spirals in active gels of polar
  filaments},\ }\href {https://doi.org/10.1103/physrevlett.92.078101}
  {\bibfield  {journal} {\bibinfo  {journal} {Physical Review Letters}\
  }\textbf {\bibinfo {volume} {92}},\ \bibinfo {pages} {078101} (\bibinfo
  {year} {2004})}\BibitemShut {NoStop}%
\bibitem [{\citenamefont {Guirao}\ and\ \citenamefont
  {Joanny}(2007)}]{Guirao2007}%
  \BibitemOpen
  \bibfield  {author} {\bibinfo {author} {\bibfnamefont {B.}~\bibnamefont
  {Guirao}}\ and\ \bibinfo {author} {\bibfnamefont {J.-F.}\ \bibnamefont
  {Joanny}},\ }\bibfield  {title} {\bibinfo {title} {Spontaneous creation of
  macroscopic flow and metachronal waves in an array of cilia},\ }\href
  {https://doi.org/10.1529/biophysj.106.084897} {\bibfield  {journal} {\bibinfo
   {journal} {Biophysical Journal}\ }\textbf {\bibinfo {volume} {92}},\
  \bibinfo {pages} {1900–1917} (\bibinfo {year} {2007})}\BibitemShut
  {NoStop}%
\bibitem [{\citenamefont {Chen}\ \emph {et~al.}(2025)\citenamefont {Chen},
  \citenamefont {Fan}, \citenamefont {Fischer}, \citenamefont {Ghosh},
  \citenamefont {G{\"o}pfrich}, \citenamefont {Golestanian}, \citenamefont
  {Hess}, \citenamefont {Ma}, \citenamefont {Nelson}, \citenamefont
  {Pati{\~n}o~Padial}, \citenamefont {Tang}, \citenamefont {Villa},
  \citenamefont {Wang}, \citenamefont {Zhang}, \citenamefont {Sen},\ and\
  \citenamefont {S{\'a}nchez}}]{Chen2025}%
  \BibitemOpen
  \bibfield  {author} {\bibinfo {author} {\bibfnamefont {S.}~\bibnamefont
  {Chen}}, \bibinfo {author} {\bibfnamefont {D.~E.}\ \bibnamefont {Fan}},
  \bibinfo {author} {\bibfnamefont {P.}~\bibnamefont {Fischer}}, \bibinfo
  {author} {\bibfnamefont {A.}~\bibnamefont {Ghosh}}, \bibinfo {author}
  {\bibfnamefont {K.}~\bibnamefont {G{\"o}pfrich}}, \bibinfo {author}
  {\bibfnamefont {R.}~\bibnamefont {Golestanian}}, \bibinfo {author}
  {\bibfnamefont {H.}~\bibnamefont {Hess}}, \bibinfo {author} {\bibfnamefont
  {X.}~\bibnamefont {Ma}}, \bibinfo {author} {\bibfnamefont {B.~J.}\
  \bibnamefont {Nelson}}, \bibinfo {author} {\bibfnamefont {T.}~\bibnamefont
  {Pati{\~n}o~Padial}}, \bibinfo {author} {\bibfnamefont {J.}~\bibnamefont
  {Tang}}, \bibinfo {author} {\bibfnamefont {K.}~\bibnamefont {Villa}},
  \bibinfo {author} {\bibfnamefont {W.}~\bibnamefont {Wang}}, \bibinfo {author}
  {\bibfnamefont {L.}~\bibnamefont {Zhang}}, \bibinfo {author} {\bibfnamefont
  {A.}~\bibnamefont {Sen}},\ and\ \bibinfo {author} {\bibfnamefont
  {S.}~\bibnamefont {S{\'a}nchez}},\ }\bibfield  {title} {\bibinfo {title} {A
  roadmap for next-generation nanomotors},\ }\href
  {https://doi.org/10.1038/s41565-025-01962-9} {\bibfield  {journal} {\bibinfo
  {journal} {Nat. Nanotechnol.}\ }\textbf {\bibinfo {volume} {20}},\ \bibinfo
  {pages} {990} (\bibinfo {year} {2025})}\BibitemShut {NoStop}%
\bibitem [{\citenamefont {Shi}\ \emph {et~al.}(2022)\citenamefont {Shi},
  \citenamefont {Pumm}, \citenamefont {Isensee}, \citenamefont {Zhao},
  \citenamefont {Verschueren}, \citenamefont {Martin-Gonzalez}, \citenamefont
  {Golestanian}, \citenamefont {Dietz},\ and\ \citenamefont
  {Dekker}}]{Shi2022}%
  \BibitemOpen
  \bibfield  {author} {\bibinfo {author} {\bibfnamefont {X.}~\bibnamefont
  {Shi}}, \bibinfo {author} {\bibfnamefont {A.-K.}\ \bibnamefont {Pumm}},
  \bibinfo {author} {\bibfnamefont {J.}~\bibnamefont {Isensee}}, \bibinfo
  {author} {\bibfnamefont {W.}~\bibnamefont {Zhao}}, \bibinfo {author}
  {\bibfnamefont {D.}~\bibnamefont {Verschueren}}, \bibinfo {author}
  {\bibfnamefont {A.}~\bibnamefont {Martin-Gonzalez}}, \bibinfo {author}
  {\bibfnamefont {R.}~\bibnamefont {Golestanian}}, \bibinfo {author}
  {\bibfnamefont {H.}~\bibnamefont {Dietz}},\ and\ \bibinfo {author}
  {\bibfnamefont {C.}~\bibnamefont {Dekker}},\ }\bibfield  {title} {\bibinfo
  {title} {Sustained unidirectional rotation of a self-organized dna rotor on a
  nanopore},\ }\href {https://doi.org/10.1038/s41567-022-01683-z} {\bibfield
  {journal} {\bibinfo  {journal} {Nat. Phys.}\ }\textbf {\bibinfo {volume}
  {18}},\ \bibinfo {pages} {1105} (\bibinfo {year} {2022})}\BibitemShut
  {NoStop}%
\bibitem [{\citenamefont {Butler}\ \emph {et~al.}(2026)\citenamefont {Butler},
  \citenamefont {Walker}, \citenamefont {Montenegro-Johnson},\ and\
  \citenamefont {Katsamba}}]{Butler2026}%
  \BibitemOpen
  \bibfield  {author} {\bibinfo {author} {\bibfnamefont {M.~D.}\ \bibnamefont
  {Butler}}, \bibinfo {author} {\bibfnamefont {B.~J.}\ \bibnamefont {Walker}},
  \bibinfo {author} {\bibfnamefont {T.~D.}\ \bibnamefont
  {Montenegro-Johnson}},\ and\ \bibinfo {author} {\bibfnamefont
  {P.}~\bibnamefont {Katsamba}},\ }\bibfield  {title} {\bibinfo {title}
  {Elastohydrodynamics of three-dimensional chemically active filaments},\
  }\href {https://doi.org/10.1017/jfm.2026.11208} {\bibfield  {journal}
  {\bibinfo  {journal} {Journal of Fluid Mechanics}\ }\textbf {\bibinfo
  {volume} {1029}},\ \bibinfo {pages} {11208} (\bibinfo {year}
  {2026})}\BibitemShut {NoStop}%
\bibitem [{\citenamefont {Zwicker}\ \emph {et~al.}(2016)\citenamefont
  {Zwicker}, \citenamefont {Seyboldt}, \citenamefont {Weber}, \citenamefont
  {Hyman},\ and\ \citenamefont {J\"{u}licher}}]{Zwicker2016}%
  \BibitemOpen
  \bibfield  {author} {\bibinfo {author} {\bibfnamefont {D.}~\bibnamefont
  {Zwicker}}, \bibinfo {author} {\bibfnamefont {R.}~\bibnamefont {Seyboldt}},
  \bibinfo {author} {\bibfnamefont {C.~A.}\ \bibnamefont {Weber}}, \bibinfo
  {author} {\bibfnamefont {A.~A.}\ \bibnamefont {Hyman}},\ and\ \bibinfo
  {author} {\bibfnamefont {F.}~\bibnamefont {J\"{u}licher}},\ }\bibfield
  {title} {\bibinfo {title} {Growth and division of active droplets provides a
  model for protocells},\ }\href {https://doi.org/10.1038/nphys3984} {\bibfield
   {journal} {\bibinfo  {journal} {Nature Physics}\ }\textbf {\bibinfo {volume}
  {13}},\ \bibinfo {pages} {408–413} (\bibinfo {year} {2016})}\BibitemShut
  {NoStop}%
\bibitem [{\citenamefont {Golestanian}(2016)}]{Golestanian2016nphys}%
  \BibitemOpen
  \bibfield  {author} {\bibinfo {author} {\bibfnamefont {R.}~\bibnamefont
  {Golestanian}},\ }\bibfield  {title} {\bibinfo {title} {Division for
  multiplication},\ }\href {https://doi.org/10.1038/nphys3998} {\bibfield
  {journal} {\bibinfo  {journal} {Nature Physics}\ }\textbf {\bibinfo {volume}
  {13}},\ \bibinfo {pages} {323–324} (\bibinfo {year} {2016})}\BibitemShut
  {NoStop}%
\bibitem [{\citenamefont {Polin}\ \emph {et~al.}(2009)\citenamefont {Polin},
  \citenamefont {Tuval}, \citenamefont {Drescher}, \citenamefont {Gollub},\
  and\ \citenamefont {Goldstein}}]{Polin2009}%
  \BibitemOpen
  \bibfield  {author} {\bibinfo {author} {\bibfnamefont {M.}~\bibnamefont
  {Polin}}, \bibinfo {author} {\bibfnamefont {I.}~\bibnamefont {Tuval}},
  \bibinfo {author} {\bibfnamefont {K.}~\bibnamefont {Drescher}}, \bibinfo
  {author} {\bibfnamefont {J.~P.}\ \bibnamefont {Gollub}},\ and\ \bibinfo
  {author} {\bibfnamefont {R.~E.}\ \bibnamefont {Goldstein}},\ }\bibfield
  {title} {\bibinfo {title} {Chlamydomonas swims with two “gears” in a
  eukaryotic version of run-and-tumble locomotion},\ }\href
  {https://doi.org/10.1126/science.1172667} {\bibfield  {journal} {\bibinfo
  {journal} {Science}\ }\textbf {\bibinfo {volume} {325}},\ \bibinfo {pages}
  {487–490} (\bibinfo {year} {2009})}\BibitemShut {NoStop}%
\bibitem [{\citenamefont {Bennett}\ and\ \citenamefont
  {Golestanian}(2013)}]{Bennett2013}%
  \BibitemOpen
  \bibfield  {author} {\bibinfo {author} {\bibfnamefont {R.~R.}\ \bibnamefont
  {Bennett}}\ and\ \bibinfo {author} {\bibfnamefont {R.}~\bibnamefont
  {Golestanian}},\ }\bibfield  {title} {\bibinfo {title} {Emergent
  run-and-tumble behavior in a simple model of chlamydomonas with intrinsic
  noise},\ }\href {https://doi.org/10.1103/PhysRevLett.110.148102} {\bibfield
  {journal} {\bibinfo  {journal} {Phys. Rev. Lett.}\ }\textbf {\bibinfo
  {volume} {110}},\ \bibinfo {eid} {148102} (\bibinfo {year}
  {2013})}\BibitemShut {NoStop}%
\bibitem [{\citenamefont {Soto}\ and\ \citenamefont
  {Golestanian}(2015)}]{Soto2015PRE}%
  \BibitemOpen
  \bibfield  {author} {\bibinfo {author} {\bibfnamefont {R.}~\bibnamefont
  {Soto}}\ and\ \bibinfo {author} {\bibfnamefont {R.}~\bibnamefont
  {Golestanian}},\ }\bibfield  {title} {\bibinfo {title} {Self-assembly of
  active colloidal molecules with dynamic function},\ }\href
  {https://doi.org/10.1103/PhysRevE.91.052304} {\bibfield  {journal} {\bibinfo
  {journal} {Phys. Rev. E}\ }\textbf {\bibinfo {volume} {91}},\ \bibinfo
  {pages} {052304} (\bibinfo {year} {2015})}\BibitemShut {NoStop}%
\bibitem [{\citenamefont {Bonthuis}\ and\ \citenamefont
  {Golestanian}(2014)}]{Bonthuis2014}%
  \BibitemOpen
  \bibfield  {author} {\bibinfo {author} {\bibfnamefont {D.~J.}\ \bibnamefont
  {Bonthuis}}\ and\ \bibinfo {author} {\bibfnamefont {R.}~\bibnamefont
  {Golestanian}},\ }\bibfield  {title} {\bibinfo {title} {Mechanosensitive
  channel activation by diffusio-osmotic force},\ }\href
  {https://doi.org/10.1103/PhysRevLett.113.148101} {\bibfield  {journal}
  {\bibinfo  {journal} {Phys. Rev. Lett.}\ }\textbf {\bibinfo {volume} {113}},\
  \bibinfo {eid} {148101} (\bibinfo {year} {2014})}\BibitemShut {NoStop}%
\bibitem [{\citenamefont {Soto}\ and\ \citenamefont
  {Golestanian}(2014)}]{Soto2014PRL}%
  \BibitemOpen
  \bibfield  {author} {\bibinfo {author} {\bibfnamefont {R.}~\bibnamefont
  {Soto}}\ and\ \bibinfo {author} {\bibfnamefont {R.}~\bibnamefont
  {Golestanian}},\ }\bibfield  {title} {\bibinfo {title} {Self-{{Assembly}} of
  {{Catalytically Active Colloidal Molecules}}: {{Tailoring Activity Through
  Surface Chemistry}}},\ }\href
  {https://doi.org/10.1103/PhysRevLett.112.068301} {\bibfield  {journal}
  {\bibinfo  {journal} {Physical Review Letters}\ }\textbf {\bibinfo {volume}
  {112}},\ \bibinfo {pages} {068301} (\bibinfo {year} {2014})}\BibitemShut
  {NoStop}%
\bibitem [{\citenamefont {Ivlev}\ \emph {et~al.}(2015)\citenamefont {Ivlev},
  \citenamefont {Bartnick}, \citenamefont {Heinen}, \citenamefont {Du},
  \citenamefont {Nosenko},\ and\ \citenamefont {L{\"o}wen}}]{Ivlev2015PRX}%
  \BibitemOpen
  \bibfield  {author} {\bibinfo {author} {\bibfnamefont {A.~V.}\ \bibnamefont
  {Ivlev}}, \bibinfo {author} {\bibfnamefont {J.}~\bibnamefont {Bartnick}},
  \bibinfo {author} {\bibfnamefont {M.}~\bibnamefont {Heinen}}, \bibinfo
  {author} {\bibfnamefont {C.-R.}\ \bibnamefont {Du}}, \bibinfo {author}
  {\bibfnamefont {V.}~\bibnamefont {Nosenko}},\ and\ \bibinfo {author}
  {\bibfnamefont {H.}~\bibnamefont {L{\"o}wen}},\ }\bibfield  {title} {\bibinfo
  {title} {Statistical {{Mechanics}} where {{Newton}}'s {{Third Law}} is
  {{Broken}}},\ }\href {https://doi.org/10.1103/PhysRevX.5.011035} {\bibfield
  {journal} {\bibinfo  {journal} {Physical Review X}\ }\textbf {\bibinfo
  {volume} {5}},\ \bibinfo {pages} {011035} (\bibinfo {year}
  {2015})}\BibitemShut {NoStop}%
\bibitem [{\citenamefont {Saha}\ \emph {et~al.}(2020)\citenamefont {Saha},
  \citenamefont {Agudo-Canalejo},\ and\ \citenamefont
  {Golestanian}}]{Saha2020}%
  \BibitemOpen
  \bibfield  {author} {\bibinfo {author} {\bibfnamefont {S.}~\bibnamefont
  {Saha}}, \bibinfo {author} {\bibfnamefont {J.}~\bibnamefont
  {Agudo-Canalejo}},\ and\ \bibinfo {author} {\bibfnamefont {R.}~\bibnamefont
  {Golestanian}},\ }\bibfield  {title} {\bibinfo {title} {Scalar active
  mixtures: The nonreciprocal cahn-hilliard model},\ }\href
  {https://doi.org/10.1103/PhysRevX.10.041009} {\bibfield  {journal} {\bibinfo
  {journal} {Phys. Rev. X}\ }\textbf {\bibinfo {volume} {10}},\ \bibinfo {eid}
  {041009} (\bibinfo {year} {2020})}\BibitemShut {NoStop}%
\bibitem [{\citenamefont {You}\ \emph {et~al.}(2020)\citenamefont {You},
  \citenamefont {Baskaran},\ and\ \citenamefont {Marchetti}}]{You2020PNAS}%
  \BibitemOpen
  \bibfield  {author} {\bibinfo {author} {\bibfnamefont {Z.}~\bibnamefont
  {You}}, \bibinfo {author} {\bibfnamefont {A.}~\bibnamefont {Baskaran}},\ and\
  \bibinfo {author} {\bibfnamefont {M.~C.}\ \bibnamefont {Marchetti}},\
  }\bibfield  {title} {\bibinfo {title} {Nonreciprocity as a generic route to
  traveling states},\ }\href {https://doi.org/10.1073/pnas.2010318117}
  {\bibfield  {journal} {\bibinfo  {journal} {Proceedings of the National
  Academy of Sciences}\ }\textbf {\bibinfo {volume} {117}},\ \bibinfo {pages}
  {19767} (\bibinfo {year} {2020})}\BibitemShut {NoStop}%
\bibitem [{\citenamefont {Fruchart}\ \emph {et~al.}(2021)\citenamefont
  {Fruchart}, \citenamefont {Hanai}, \citenamefont {Littlewood},\ and\
  \citenamefont {Vitelli}}]{Fruchart2021N}%
  \BibitemOpen
  \bibfield  {author} {\bibinfo {author} {\bibfnamefont {M.}~\bibnamefont
  {Fruchart}}, \bibinfo {author} {\bibfnamefont {R.}~\bibnamefont {Hanai}},
  \bibinfo {author} {\bibfnamefont {P.~B.}\ \bibnamefont {Littlewood}},\ and\
  \bibinfo {author} {\bibfnamefont {V.}~\bibnamefont {Vitelli}},\ }\bibfield
  {title} {\bibinfo {title} {Non-reciprocal phase transitions},\ }\href
  {https://doi.org/10.1038/s41586-021-03375-9} {\bibfield  {journal} {\bibinfo
  {journal} {Nature}\ }\textbf {\bibinfo {volume} {592}},\ \bibinfo {pages}
  {363} (\bibinfo {year} {2021})}\BibitemShut {NoStop}%
\bibitem [{\citenamefont {Kreienkamp}\ and\ \citenamefont
  {Klapp}(2022)}]{Kreienkamp2022NJP}%
  \BibitemOpen
  \bibfield  {author} {\bibinfo {author} {\bibfnamefont {K.~L.}\ \bibnamefont
  {Kreienkamp}}\ and\ \bibinfo {author} {\bibfnamefont {S.~H.~L.}\ \bibnamefont
  {Klapp}},\ }\bibfield  {title} {\bibinfo {title} {Clustering and flocking of
  repulsive chiral active particles with non-reciprocal couplings},\ }\href
  {https://doi.org/10.1088/1367-2630/ac9cc3} {\bibfield  {journal} {\bibinfo
  {journal} {New Journal of Physics}\ }\textbf {\bibinfo {volume} {24}},\
  \bibinfo {pages} {123009} (\bibinfo {year} {2022})}\BibitemShut {NoStop}%
\bibitem [{\citenamefont {Dinelli}\ \emph {et~al.}(2023)\citenamefont
  {Dinelli}, \citenamefont {O'Byrne}, \citenamefont {Curatolo}, \citenamefont
  {Zhao}, \citenamefont {Sollich},\ and\ \citenamefont
  {Tailleur}}]{Dinelli2023NC}%
  \BibitemOpen
  \bibfield  {author} {\bibinfo {author} {\bibfnamefont {A.}~\bibnamefont
  {Dinelli}}, \bibinfo {author} {\bibfnamefont {J.}~\bibnamefont {O'Byrne}},
  \bibinfo {author} {\bibfnamefont {A.}~\bibnamefont {Curatolo}}, \bibinfo
  {author} {\bibfnamefont {Y.}~\bibnamefont {Zhao}}, \bibinfo {author}
  {\bibfnamefont {P.}~\bibnamefont {Sollich}},\ and\ \bibinfo {author}
  {\bibfnamefont {J.}~\bibnamefont {Tailleur}},\ }\bibfield  {title} {\bibinfo
  {title} {Non-reciprocity across scales in active mixtures},\ }\href
  {https://doi.org/10.1038/s41467-023-42713-5} {\bibfield  {journal} {\bibinfo
  {journal} {Nature Communications}\ }\textbf {\bibinfo {volume} {14}},\
  \bibinfo {pages} {7035} (\bibinfo {year} {2023})}\BibitemShut {NoStop}%
\bibitem [{\citenamefont {Osat}\ and\ \citenamefont
  {Golestanian}(2022)}]{Osat2022}%
  \BibitemOpen
  \bibfield  {author} {\bibinfo {author} {\bibfnamefont {S.}~\bibnamefont
  {Osat}}\ and\ \bibinfo {author} {\bibfnamefont {R.}~\bibnamefont
  {Golestanian}},\ }\bibfield  {title} {\bibinfo {title} {Non-reciprocal
  multifarious self-organization},\ }\href
  {https://doi.org/10.1038/s41565-022-01258-2} {\bibfield  {journal} {\bibinfo
  {journal} {Nat. Nanotechnol.}\ }\textbf {\bibinfo {volume} {18}},\ \bibinfo
  {pages} {79} (\bibinfo {year} {2022})}\BibitemShut {NoStop}%
\bibitem [{\citenamefont {Ouazan-Reboul}\ \emph {et~al.}(2023)\citenamefont
  {Ouazan-Reboul}, \citenamefont {Agudo-Canalejo},\ and\ \citenamefont
  {Golestanian}}]{OuazanReboul2023b}%
  \BibitemOpen
  \bibfield  {author} {\bibinfo {author} {\bibfnamefont {V.}~\bibnamefont
  {Ouazan-Reboul}}, \bibinfo {author} {\bibfnamefont {J.}~\bibnamefont
  {Agudo-Canalejo}},\ and\ \bibinfo {author} {\bibfnamefont {R.}~\bibnamefont
  {Golestanian}},\ }\bibfield  {title} {\bibinfo {title} {Self-organization of
  primitive metabolic cycles due to non-reciprocal interactions},\ }\href
  {https://doi.org/10.1038/s41467-023-40241-w} {\bibfield  {journal} {\bibinfo
  {journal} {Nat. Commun.}\ }\textbf {\bibinfo {volume} {14}},\ \bibinfo {eid}
  {4496} (\bibinfo {year} {2023})}\BibitemShut {NoStop}%
\bibitem [{\citenamefont {{Agudo-Canalejo}}\ and\ \citenamefont
  {Golestanian}(2019)}]{Agudo-Canalejo2019PRL}%
  \BibitemOpen
  \bibfield  {author} {\bibinfo {author} {\bibfnamefont {J.}~\bibnamefont
  {{Agudo-Canalejo}}}\ and\ \bibinfo {author} {\bibfnamefont {R.}~\bibnamefont
  {Golestanian}},\ }\bibfield  {title} {\bibinfo {title} {Active {{Phase
  Separation}} in {{Mixtures}} of {{Chemically Interacting Particles}}},\
  }\href {https://doi.org/10.1103/PhysRevLett.123.018101} {\bibfield  {journal}
  {\bibinfo  {journal} {Physical Review Letters}\ }\textbf {\bibinfo {volume}
  {123}},\ \bibinfo {pages} {018101} (\bibinfo {year} {2019})}\BibitemShut
  {NoStop}%
\bibitem [{\citenamefont {Saha}\ \emph {et~al.}(2019)\citenamefont {Saha},
  \citenamefont {Ramaswamy},\ and\ \citenamefont {Golestanian}}]{Saha2019NJP}%
  \BibitemOpen
  \bibfield  {author} {\bibinfo {author} {\bibfnamefont {S.}~\bibnamefont
  {Saha}}, \bibinfo {author} {\bibfnamefont {S.}~\bibnamefont {Ramaswamy}},\
  and\ \bibinfo {author} {\bibfnamefont {R.}~\bibnamefont {Golestanian}},\
  }\bibfield  {title} {\bibinfo {title} {Pairing, waltzing and scattering of
  chemotactic active colloids},\ }\href
  {https://doi.org/10.1088/1367-2630/ab20fd} {\bibfield  {journal} {\bibinfo
  {journal} {New Journal of Physics}\ }\textbf {\bibinfo {volume} {21}},\
  \bibinfo {pages} {063006} (\bibinfo {year} {2019})}\BibitemShut {NoStop}%
\bibitem [{\citenamefont {Gupta}\ \emph {et~al.}(2022)\citenamefont {Gupta},
  \citenamefont {Kant}, \citenamefont {Soni}, \citenamefont {Sood},\ and\
  \citenamefont {Ramaswamy}}]{Gupta2022PRE}%
  \BibitemOpen
  \bibfield  {author} {\bibinfo {author} {\bibfnamefont {R.~K.}\ \bibnamefont
  {Gupta}}, \bibinfo {author} {\bibfnamefont {R.}~\bibnamefont {Kant}},
  \bibinfo {author} {\bibfnamefont {H.}~\bibnamefont {Soni}}, \bibinfo {author}
  {\bibfnamefont {A.~K.}\ \bibnamefont {Sood}},\ and\ \bibinfo {author}
  {\bibfnamefont {S.}~\bibnamefont {Ramaswamy}},\ }\bibfield  {title} {\bibinfo
  {title} {Active nonreciprocal attraction between motile particles in an
  elastic medium},\ }\href {https://doi.org/10.1103/PhysRevE.105.064602}
  {\bibfield  {journal} {\bibinfo  {journal} {Physical Review E}\ }\textbf
  {\bibinfo {volume} {105}},\ \bibinfo {pages} {064602} (\bibinfo {year}
  {2022})}\BibitemShut {NoStop}%
\bibitem [{\citenamefont {Lavergne}\ \emph {et~al.}(2019)\citenamefont
  {Lavergne}, \citenamefont {Wendehenne}, \citenamefont {B{\"a}uerle},\ and\
  \citenamefont {Bechinger}}]{Lavergne2019S}%
  \BibitemOpen
  \bibfield  {author} {\bibinfo {author} {\bibfnamefont {F.~A.}\ \bibnamefont
  {Lavergne}}, \bibinfo {author} {\bibfnamefont {H.}~\bibnamefont
  {Wendehenne}}, \bibinfo {author} {\bibfnamefont {T.}~\bibnamefont
  {B{\"a}uerle}},\ and\ \bibinfo {author} {\bibfnamefont {C.}~\bibnamefont
  {Bechinger}},\ }\bibfield  {title} {\bibinfo {title} {Group formation and
  cohesion of active particles with visual perception--dependent motility},\
  }\href {https://doi.org/10.1126/science.aau5347} {\bibfield  {journal}
  {\bibinfo  {journal} {Science}\ }\textbf {\bibinfo {volume} {364}},\ \bibinfo
  {pages} {70} (\bibinfo {year} {2019})}\BibitemShut {NoStop}%
\bibitem [{\citenamefont {Borges-Fernandes}\ \emph {et~al.}(2025)\citenamefont
  {Borges-Fernandes}, \citenamefont {Apriceno}, \citenamefont
  {Arango-Restrepo}, \citenamefont {Almadhi}, \citenamefont {Ghosh},
  \citenamefont {Forth}, \citenamefont {López-Alonso}, \citenamefont
  {Ubarretxena-Belandia}, \citenamefont {Rubi}, \citenamefont {Ruiz-Pérez},
  \citenamefont {Williams},\ and\ \citenamefont
  {Battaglia}}]{BorgesFernandes2025}%
  \BibitemOpen
  \bibfield  {author} {\bibinfo {author} {\bibfnamefont {B.}~\bibnamefont
  {Borges-Fernandes}}, \bibinfo {author} {\bibfnamefont {A.}~\bibnamefont
  {Apriceno}}, \bibinfo {author} {\bibfnamefont {A.}~\bibnamefont
  {Arango-Restrepo}}, \bibinfo {author} {\bibfnamefont {S.}~\bibnamefont
  {Almadhi}}, \bibinfo {author} {\bibfnamefont {S.}~\bibnamefont {Ghosh}},
  \bibinfo {author} {\bibfnamefont {J.}~\bibnamefont {Forth}}, \bibinfo
  {author} {\bibfnamefont {J.~P.}\ \bibnamefont {López-Alonso}}, \bibinfo
  {author} {\bibfnamefont {I.}~\bibnamefont {Ubarretxena-Belandia}}, \bibinfo
  {author} {\bibfnamefont {J.~M.}\ \bibnamefont {Rubi}}, \bibinfo {author}
  {\bibfnamefont {L.}~\bibnamefont {Ruiz-Pérez}}, \bibinfo {author}
  {\bibfnamefont {I.}~\bibnamefont {Williams}},\ and\ \bibinfo {author}
  {\bibfnamefont {G.}~\bibnamefont {Battaglia}},\ }\bibfield  {title} {\bibinfo
  {title} {The minimal chemotactic cell},\ }\bibfield  {journal} {\bibinfo
  {journal} {Science Advances}\ }\textbf {\bibinfo {volume} {11}},\ \href
  {https://doi.org/10.1126/sciadv.adx9364} {10.1126/sciadv.adx9364} (\bibinfo
  {year} {2025})\BibitemShut {NoStop}%
\bibitem [{\citenamefont {Golestanian}(2022)}]{Golestanian2022a}%
  \BibitemOpen
  \bibfield  {author} {\bibinfo {author} {\bibfnamefont {R.}~\bibnamefont
  {Golestanian}},\ }\bibfield  {title} {\bibinfo {title} {Phoretic active
  matter},\ }in\ \href@noop {} {\emph {\bibinfo {booktitle} {{Active Matter and
  Nonequilibrium Statistical Physics. Lecture Notes of the Les Houches Summer
  School}}}}\ (\bibinfo  {publisher} {Oxford University Press},\ \bibinfo
  {address} {UK},\ \bibinfo {year} {2022})\ pp.\ \bibinfo {pages}
  {230--293}\BibitemShut {NoStop}%
\bibitem [{\citenamefont {Bardeen}\ \emph {et~al.}(1957)\citenamefont
  {Bardeen}, \citenamefont {Cooper},\ and\ \citenamefont
  {Schrieffer}}]{BCS1957}%
  \BibitemOpen
  \bibfield  {author} {\bibinfo {author} {\bibfnamefont {J.}~\bibnamefont
  {Bardeen}}, \bibinfo {author} {\bibfnamefont {L.~N.}\ \bibnamefont
  {Cooper}},\ and\ \bibinfo {author} {\bibfnamefont {J.~R.}\ \bibnamefont
  {Schrieffer}},\ }\bibfield  {title} {\bibinfo {title} {Theory of
  superconductivity},\ }\href {https://doi.org/10.1103/PhysRev.108.1175}
  {\bibfield  {journal} {\bibinfo  {journal} {Phys. Rev.}\ }\textbf {\bibinfo
  {volume} {108}},\ \bibinfo {pages} {1175} (\bibinfo {year}
  {1957})}\BibitemShut {NoStop}%
\bibitem [{\citenamefont {Lee}\ \emph {et~al.}(2006)\citenamefont {Lee},
  \citenamefont {Nagaosa},\ and\ \citenamefont {Wen}}]{LeeRMP2006}%
  \BibitemOpen
  \bibfield  {author} {\bibinfo {author} {\bibfnamefont {P.~A.}\ \bibnamefont
  {Lee}}, \bibinfo {author} {\bibfnamefont {N.}~\bibnamefont {Nagaosa}},\ and\
  \bibinfo {author} {\bibfnamefont {X.-G.}\ \bibnamefont {Wen}},\ }\bibfield
  {title} {\bibinfo {title} {Doping a mott insulator: Physics of
  high-temperature superconductivity},\ }\href
  {https://doi.org/10.1103/RevModPhys.78.17} {\bibfield  {journal} {\bibinfo
  {journal} {Rev. Mod. Phys.}\ }\textbf {\bibinfo {volume} {78}},\ \bibinfo
  {pages} {17} (\bibinfo {year} {2006})}\BibitemShut {NoStop}%
\bibitem [{\citenamefont {Pisegna}\ \emph {et~al.}(2024)\citenamefont
  {Pisegna}, \citenamefont {Saha},\ and\ \citenamefont
  {Golestanian}}]{Pisegna2024}%
  \BibitemOpen
  \bibfield  {author} {\bibinfo {author} {\bibfnamefont {G.}~\bibnamefont
  {Pisegna}}, \bibinfo {author} {\bibfnamefont {S.}~\bibnamefont {Saha}},\ and\
  \bibinfo {author} {\bibfnamefont {R.}~\bibnamefont {Golestanian}},\
  }\bibfield  {title} {\bibinfo {title} {Emergent polar order in nonpolar
  mixtures with nonreciprocal interactions},\ }\href
  {https://doi.org/10.1073/pnas.2407705121} {\bibfield  {journal} {\bibinfo
  {journal} {PNAS}\ }\textbf {\bibinfo {volume} {121}},\ \bibinfo {eid}
  {e2407705121} (\bibinfo {year} {2024})}\BibitemShut {NoStop}%
\bibitem [{\citenamefont {Cohen}\ and\ \citenamefont
  {Golestanian}(2014)}]{Cohen2014}%
  \BibitemOpen
  \bibfield  {author} {\bibinfo {author} {\bibfnamefont {J.~A.}\ \bibnamefont
  {Cohen}}\ and\ \bibinfo {author} {\bibfnamefont {R.}~\bibnamefont
  {Golestanian}},\ }\bibfield  {title} {\bibinfo {title} {Emergent cometlike
  swarming of optically driven thermally active colloids},\ }\href
  {https://doi.org/10.1103/PhysRevLett.112.068302} {\bibfield  {journal}
  {\bibinfo  {journal} {Phys. Rev. Lett.}\ }\textbf {\bibinfo {volume} {112}},\
  \bibinfo {eid} {068302} (\bibinfo {year} {2014})}\BibitemShut {NoStop}%
\bibitem [{\citenamefont {{van Kesteren}}\ \emph {et~al.}(2023)\citenamefont
  {{van Kesteren}}, \citenamefont {Alvarez}, \citenamefont {{Arrese-Igor}},
  \citenamefont {Alegria},\ and\ \citenamefont {Isa}}]{vanKesteren2023PNAS}%
  \BibitemOpen
  \bibfield  {author} {\bibinfo {author} {\bibfnamefont {S.}~\bibnamefont {{van
  Kesteren}}}, \bibinfo {author} {\bibfnamefont {L.}~\bibnamefont {Alvarez}},
  \bibinfo {author} {\bibfnamefont {S.}~\bibnamefont {{Arrese-Igor}}}, \bibinfo
  {author} {\bibfnamefont {A.}~\bibnamefont {Alegria}},\ and\ \bibinfo {author}
  {\bibfnamefont {L.}~\bibnamefont {Isa}},\ }\bibfield  {title} {\bibinfo
  {title} {Self-propelling colloids with finite state dynamics},\ }\href
  {https://doi.org/10.1073/pnas.2213481120} {\bibfield  {journal} {\bibinfo
  {journal} {Proceedings of the National Academy of Sciences}\ }\textbf
  {\bibinfo {volume} {120}},\ \bibinfo {pages} {e2213481120} (\bibinfo {year}
  {2023})}\BibitemShut {NoStop}%
\bibitem [{\citenamefont {Boniface}\ \emph {et~al.}(2024)\citenamefont
  {Boniface}, \citenamefont {Leyva}, \citenamefont {Pagonabarraga},\ and\
  \citenamefont {Tierno}}]{Boniface2024NC}%
  \BibitemOpen
  \bibfield  {author} {\bibinfo {author} {\bibfnamefont {D.}~\bibnamefont
  {Boniface}}, \bibinfo {author} {\bibfnamefont {S.~G.}\ \bibnamefont {Leyva}},
  \bibinfo {author} {\bibfnamefont {I.}~\bibnamefont {Pagonabarraga}},\ and\
  \bibinfo {author} {\bibfnamefont {P.}~\bibnamefont {Tierno}},\ }\bibfield
  {title} {\bibinfo {title} {Clustering induces switching between phoretic and
  osmotic propulsion in active colloidal rafts},\ }\href
  {https://doi.org/10.1038/s41467-024-49977-5} {\bibfield  {journal} {\bibinfo
  {journal} {Nature Communications}\ }\textbf {\bibinfo {volume} {15}},\
  \bibinfo {pages} {5666} (\bibinfo {year} {2024})}\BibitemShut {NoStop}%
\bibitem [{\citenamefont {Peng}\ and\ \citenamefont
  {Kapral}(2024)}]{Peng2024SM}%
  \BibitemOpen
  \bibfield  {author} {\bibinfo {author} {\bibfnamefont {Z.}~\bibnamefont
  {Peng}}\ and\ \bibinfo {author} {\bibfnamefont {R.}~\bibnamefont {Kapral}},\
  }\bibfield  {title} {\bibinfo {title} {Self-organization of active colloids
  mediated by chemical interactions},\ }\href
  {https://doi.org/10.1039/D3SM01272G} {\bibfield  {journal} {\bibinfo
  {journal} {Soft Matter}\ }\textbf {\bibinfo {volume} {20}},\ \bibinfo {pages}
  {1100} (\bibinfo {year} {2024})}\BibitemShut {NoStop}%
\bibitem [{\citenamefont {Wioland}\ \emph {et~al.}(2013)\citenamefont
  {Wioland}, \citenamefont {Woodhouse}, \citenamefont {Dunkel}, \citenamefont
  {Kessler},\ and\ \citenamefont {Goldstein}}]{Wioland2013}%
  \BibitemOpen
  \bibfield  {author} {\bibinfo {author} {\bibfnamefont {H.}~\bibnamefont
  {Wioland}}, \bibinfo {author} {\bibfnamefont {F.~G.}\ \bibnamefont
  {Woodhouse}}, \bibinfo {author} {\bibfnamefont {J.}~\bibnamefont {Dunkel}},
  \bibinfo {author} {\bibfnamefont {J.~O.}\ \bibnamefont {Kessler}},\ and\
  \bibinfo {author} {\bibfnamefont {R.~E.}\ \bibnamefont {Goldstein}},\
  }\bibfield  {title} {\bibinfo {title} {Confinement stabilizes a bacterial
  suspension into a spiral vortex},\ }\href
  {https://doi.org/10.1103/physrevlett.110.268102} {\bibfield  {journal}
  {\bibinfo  {journal} {Physical Review Letters}\ }\textbf {\bibinfo {volume}
  {110}},\ \bibinfo {pages} {268102} (\bibinfo {year} {2013})}\BibitemShut
  {NoStop}%
\bibitem [{\citenamefont {Wu}\ \emph {et~al.}(2017)\citenamefont {Wu},
  \citenamefont {Hishamunda}, \citenamefont {Chen}, \citenamefont {DeCamp},
  \citenamefont {Chang}, \citenamefont {Fernández-Nieves}, \citenamefont
  {Fraden},\ and\ \citenamefont {Dogic}}]{Wu2017}%
  \BibitemOpen
  \bibfield  {author} {\bibinfo {author} {\bibfnamefont {K.-T.}\ \bibnamefont
  {Wu}}, \bibinfo {author} {\bibfnamefont {J.~B.}\ \bibnamefont {Hishamunda}},
  \bibinfo {author} {\bibfnamefont {D.~T.~N.}\ \bibnamefont {Chen}}, \bibinfo
  {author} {\bibfnamefont {S.~J.}\ \bibnamefont {DeCamp}}, \bibinfo {author}
  {\bibfnamefont {Y.-W.}\ \bibnamefont {Chang}}, \bibinfo {author}
  {\bibfnamefont {A.}~\bibnamefont {Fernández-Nieves}}, \bibinfo {author}
  {\bibfnamefont {S.}~\bibnamefont {Fraden}},\ and\ \bibinfo {author}
  {\bibfnamefont {Z.}~\bibnamefont {Dogic}},\ }\bibfield  {title} {\bibinfo
  {title} {Transition from turbulent to coherent flows in confined
  three-dimensional active fluids},\ }\href
  {https://doi.org/10.1126/science.aal1979} {\bibfield  {journal} {\bibinfo
  {journal} {Science}\ }\textbf {\bibinfo {volume} {355}},\ \bibinfo {pages}
  {aal1979} (\bibinfo {year} {2017})}\BibitemShut {NoStop}%
\bibitem [{\citenamefont {Weeks}\ \emph {et~al.}(1971)\citenamefont {Weeks},
  \citenamefont {Chandler},\ and\ \citenamefont {Andersen}}]{Weeks1971JCP}%
  \BibitemOpen
  \bibfield  {author} {\bibinfo {author} {\bibfnamefont {J.~D.}\ \bibnamefont
  {Weeks}}, \bibinfo {author} {\bibfnamefont {D.}~\bibnamefont {Chandler}},\
  and\ \bibinfo {author} {\bibfnamefont {H.~C.}\ \bibnamefont {Andersen}},\
  }\bibfield  {title} {\bibinfo {title} {Role of {{Repulsive Forces}} in
  {{Determining}} the {{Equilibrium Structure}} of {{Simple Liquids}}},\ }\href
  {https://doi.org/10.1063/1.1674820} {\bibfield  {journal} {\bibinfo
  {journal} {The Journal of Chemical Physics}\ }\textbf {\bibinfo {volume}
  {54}},\ \bibinfo {pages} {5237} (\bibinfo {year} {1971})}\BibitemShut
  {NoStop}%
\bibitem [{\citenamefont {Golestanian}(2015)}]{Golestanian2015PRL}%
  \BibitemOpen
  \bibfield  {author} {\bibinfo {author} {\bibfnamefont {R.}~\bibnamefont
  {Golestanian}},\ }\bibfield  {title} {\bibinfo {title} {Enhanced
  {{Diffusion}} of {{Enzymes}} that {{Catalyze Exothermic Reactions}}},\ }\href
  {https://doi.org/10.1103/PhysRevLett.115.108102} {\bibfield  {journal}
  {\bibinfo  {journal} {Physical Review Letters}\ }\textbf {\bibinfo {volume}
  {115}},\ \bibinfo {pages} {108102} (\bibinfo {year} {2015})}\BibitemShut
  {NoStop}%
\end{thebibliography}%

\end{document}


\title{Emergent single-species non-reciprocity from bistable chemical dynamics \\
{\it Supplementary Information}}

\author{Jakob Metson}
\affiliation{Max Planck Institute for Dynamics and Self-Organization (MPI-DS), 37077 G\"ottingen, Germany}

\author{Ramin Golestanian}
\affiliation{Max Planck Institute for Dynamics and Self-Organization (MPI-DS), 37077 G\"ottingen, Germany}
\affiliation{Rudolf Peierls Centre for Theoretical Physics, University of Oxford, Oxford OX1 3PU, United Kingdom}


\maketitle

\tableofcontents

\clearpage

\section{Equations of motion}

\subsection{Non-dimensionalization}

To non-dimensionalize our equations, we use the particle diameter $\sigma$ as a length scale and the inverse chemical exchange coefficient $\Gamma^{-1}$ as a time scale. The friction coefficient $\zeta$ is removed by re-scaling the WCA potential scale $\epsilon$ and the confinement potential scale $k$. We rescale all concentrations by the Michaelis constant (or Hill function half-concentration) $K$. 
We furthermore use $V_d\approx \sigma^d$, neglecting a geometric prefactor of order one.
Simulations and numerical results are always presented using the dimensionless variables.

The transition to the non-dimensional form of the variables and parameters denoted with tilde henceforth is carried out as follows: 
$t \to \di{t}/\Gamma$, $\vec{r}_i \to \di{\vec{r}}_i \sigma$, $C \to K \di{C}$, $c_i \to K \di{c}_i$, $ \mu \to \di{\mu} \sigma^2 \Gamma/K$, $\Dp \to \di{\Dp} {\sigma}^{2}{\Gamma}$, $\epsilon \to \di{\epsilon} {\zeta}{\sigma}^2{\Gamma}$, ${k} \to \di{k} {\zeta}{\Gamma}$, $\Rbdr \to \di{R}_\text{bdr}{\sigma}$, ${\Dc} \to \di{\Dc} {\sigma}^2{\Gamma}$, ${\gamma} \to \di{\gamma} {\Gamma}$, ${s} \to \di{s} {\Gamma} $, ${a} \to \di{a} \Gamma$, ${b} \to \di{b}{\Gamma}$, and ${Q} \to \di{Q} K {\Gamma}$. We then remove the tilde when reporting the simulations and numerical results for simplicity.


\subsection{Setting parameters}

Our results are not strongly affected by the WCA potential scale $\epsilon$, and as such we fix the dimensionless $\epsilon=1$. This scale keeps the repulsion sufficiently strong such that the particles maintain a separation of approximately their diameter when touching, whilst preventing numerical instabilities associated with an excessively strong repulsion.
Furthermore, the particle diffusion coefficient $\Dp$ also does not strongly influence the results, provided it does not overpower the chemical interactions.
Therefore, in cases where we include positional noise we fix $\Dp\approx 10^{-4}\Dc - 10^{-3}\Dc$.
The dimensionless $\mu$ sets the scale for the colloids' response to chemical gradients. We fix $|\mu|=10^5$ in dimensionless units, corresponding to the particles responding over similar scales to the chemical dynamics.

\section{Space-free chemical dynamics}
As derived in the Methods section, the space-free chemical dynamics are given by
\begin{align}
    \frac{\dif c_\O}{\dif t} &= s K + \Gamma\sum_{i} (c_i-c_\O) - \gamma c_\O \\
    \frac{\dif c_i}{\dif t}  &= f(c_i) + \Gamma(c_\O-c_i),
\end{align}
where $i$ labels $N$ different colloids.

\subsection{Single-colloid stability analysis}

With one colloid (labelled $\A$), and the background, the fixed-points are given by
\begin{align}
    0 &= f(\sst{c}_\A) + \frac{sK\,\Gamma}{\Gamma +\gamma} - \frac{\gamma\Gamma}{\Gamma +\gamma}\sst{c}_\A, \\
    \sst{c}_\O &= \frac{sK}{\Gamma+\gamma} + \frac{\Gamma}{\Gamma +\gamma}\sst{c}_\A. \label{eq:cubic:co}
\end{align}
%
We then can recast the resulting cubic equation into the depressed cubic form, using the variable transform
\begin{equation}\label{eq:cubic:xt}
    \sst{c}_\A = u K+\frac{K}{3}\left(\frac{a+\frac{s\Gamma}{\Gamma+\gamma}}{b+\frac{\Gamma \gamma}{\Gamma+\gamma}}\right),
\end{equation} 
to obtain
\begin{equation}
    u^3 + 3 pu + 2 q = 0,
\end{equation}
where in our case
\begin{align}
    p &=\frac{1}{3}-\frac{1}{9}\left(\frac{a+\frac{s\Gamma}{\Gamma+\gamma}}{b+\frac{\gamma}{\Gamma+\gamma}}\right)^2,\\
    q &=\frac{1}{2} \frac{1}{\left(b+\frac{\Gamma\gamma}{\Gamma+\gamma}\right)}\left[\frac{1}{3}\left(a+\frac{s\Gamma}{\Gamma+\gamma}\right)+\frac{s\Gamma}{\Gamma+\gamma}\right] -\frac{1}{27}\left(\frac{a+\frac{s\Gamma}{\Gamma+\gamma}}{b+\frac{\Gamma\gamma}{\Gamma+\gamma}}\right)^3.
\end{align}
We can use the discriminant of the cubic equation
\begin{align}
\Delta =& -108 (p^3+q^2) \\
\begin{split}
=& 18 \left(b+\frac{\gamma}{\Gamma+\gamma}\right)^2\left(a+\frac{s\Gamma}{\Gamma+\gamma}\right)\frac{s\Gamma}{\Gamma+\gamma} - 4\left(a+\frac{s\Gamma}{\Gamma+\gamma}\right)^3\frac{s\Gamma}{\Gamma+\gamma} \\
    &+ \left(b+\frac{\gamma}{\Gamma+\gamma}\right)^2\left(a+\frac{s\Gamma}{\Gamma+\gamma}\right)^2 - 4 \left(b+\frac{\gamma}{\Gamma+\gamma}\right)^4 - 27\left(b+\frac{\gamma}{\Gamma+\gamma}\right)^2\left(\frac{s\Gamma}{\Gamma+\gamma}\right)^2.
\end{split}
\end{align}
to identify the number of roots. When $\Delta<0$ we have one fixed point, and when $\Delta>0$ we have three fixed points.
Since this is a cubic polynomial, the roots, corresponding to $\sst{c}_\A$, can be obtained analytically
\begin{align}
    u_1 &= \sqrt[3]{-q+\sqrt{q^2+p^3}} + \sqrt[3]{-q-\sqrt{q^2+p^3}} \\
    u_2 &= \frac{-1+i\sqrt{3}}{2} u_1\\
    u_3 &= \frac{-1-i\sqrt{3}}{2} u_1.
\end{align}
Here $\sqrt{}$ and $\sqrt[3]{}$ represent the principal values of the root functions. The values of $\sst{c}_\A$ corresponding to each $u_i$ solution can be obtained using Eq.~\eqref{eq:cubic:xt}. Then the corresponding $\sst{c}_\O$ solutions are given straightforwardly by Eq.~\eqref{eq:cubic:co}.

The stability of each steady-state solution is evaluated by finding the eigenvalues of the Jacobian
\begin{align}
    \mat{J}(\{\sst{c}_i\}) = \begin{pmatrix}
-\Gamma - \gamma & \Gamma \\
\Gamma & f'(\sst{c}_\A)-\Gamma
\end{pmatrix},
\end{align}
where $f'(\sst{c}_\A) = \frac{\dif{f}}{\dif{c}}|_{\sst{c}_\A}$. The steady-state is stable if all the eigenvalues are negative.

\subsection{Two-colloid stability analysis}

With two colloids (labelled $\A$ and $\B$), and the background, the fixed-point equations are given as
\begin{align}
    &s K + \Gamma (\sst{c}_\A-\sst{c}_\O)+\Gamma (\sst{c}_\B-\sst{c}_\O)- \gamma\sst{c}_\O=0, \label{eq:2c0-1} \\ 
    &f(\sst{c}_\A)+\Gamma (\sst{c}_\O-\sst{c}_\A)=0, \label{eq:2c0-2}\\
    &f(\sst{c}_\B)+\Gamma (\sst{c}_\O-\sst{c}_\B)=0. \label{eq:2c0-3}
\end{align}
which can be rewritten as follows
\begin{align}
    &f\bigg( \left(1 + \gamma/\Gamma\right)\sst{c}_\A-(2 + \gamma/\Gamma)f(\sst{c}_\A)/\Gamma-s K/\Gamma \bigg) + (1 + \gamma/\Gamma)f(\sst{c}_\A) + s K - \gamma\sst{c}_\A=0, \label{eq:2c-1} \\ 
    &\sst{c}_\B = (1 + \gamma/\Gamma)\sst{c}_\A-(2 + \gamma/\Gamma)f(\sst{c}_\A)/\Gamma-s K/\Gamma, \label{eq:2c-2}\\
    &\sst{c}_\O = \frac{S}{2\Gamma +\gamma} + \frac{\Gamma}{2\Gamma +\gamma}(\sst{c}_\A+\sst{c}_\B). \label{eq:2c-3}
\end{align}
Equation \eqref{eq:2c-1} can be recast in the form of a polynomial (we use \texttt{SymPy} \cite{SymPy} to find the coefficients).
The roots of the polynomial (solutions for $\sst{c}_\A$) are found numerically, and the corresponding $\sst{c}_\B$ and $\sst{c}_\O$ values are found using Eqs. \eqref{eq:2c-2} and \eqref{eq:2c-3}.

The eigenvalues of the Jacobian
\begin{align}
    \mat{J}(\{\sst{c}_i\}) = \begin{pmatrix}
-\Gamma - \gamma & \Gamma & \Gamma \\
\Gamma & f'(\sst{c}_\A)-\Gamma & 0 \\
\Gamma & 0 & f'(\sst{c}_\B)-\Gamma \\
\end{pmatrix}
\end{align}
are found numerically using \texttt{NumPy} \cite{NumPy} and again determine the stability of each fixed point.

With multiple ($N>2$) colloids, the fixed-points can in principle be found in a similar way. However, with multiple colloids, it may become more straightforward to find the solutions numerically directly from the original fixed-point equations.

\begin{figure}[t]
\centering
\includegraphics[width=\linewidth]{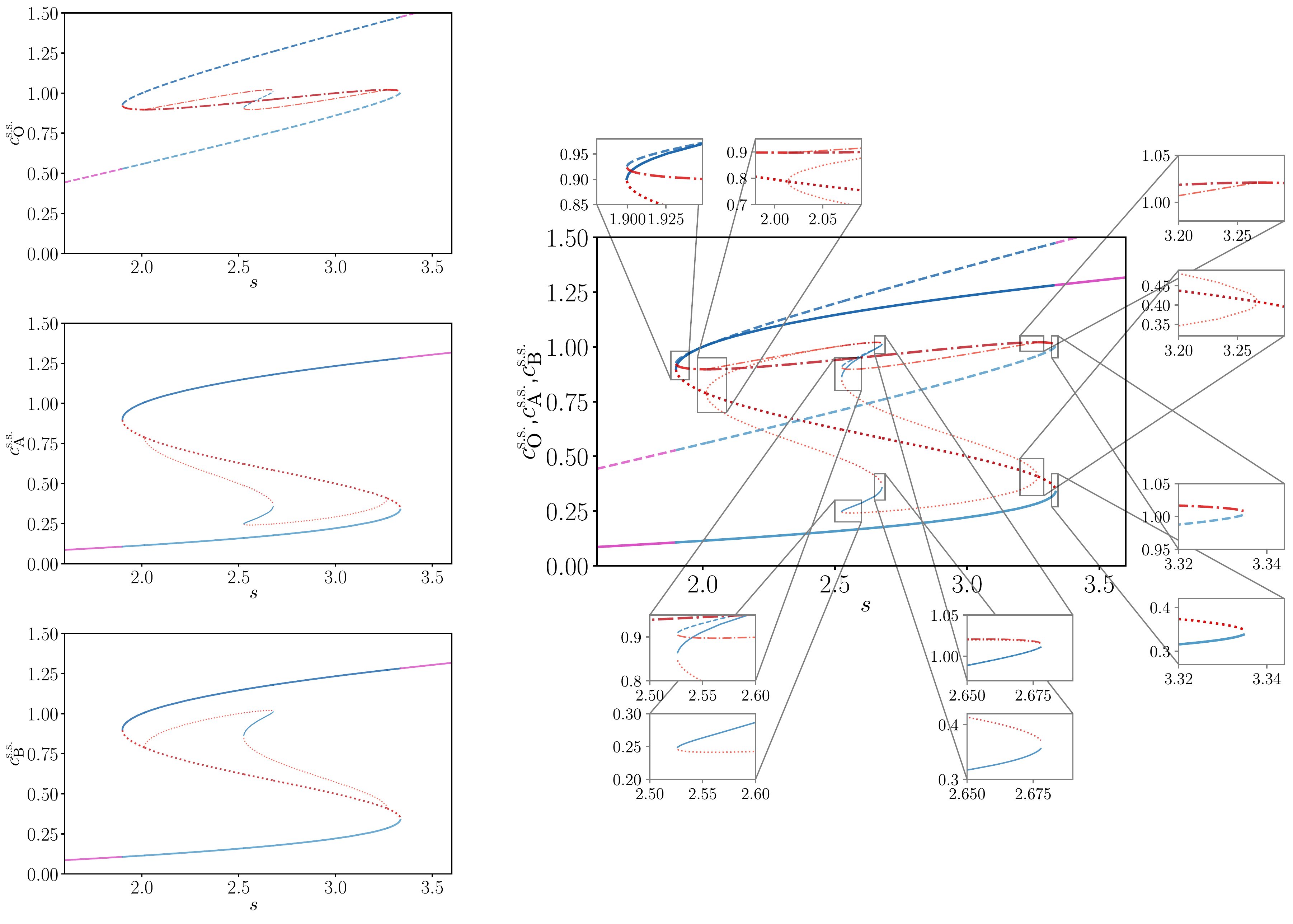}
\caption{\textbf{Bifurcation diagram of the chemical dynamics.} 
Fixed-points of the space-free chemical dynamics as $s$ is varied, with $a=10,b=5,\gamma=2$. Left: separate versions of the diagram for each concentration. Right: a version that combines the three concentrations with close-ups on each of the bifurcations. 
Solid lines show the value of $\sst{c}_\A$ and $\sst{c}_\B$ at stable fixed-points, with dashed lines indicating the corresponding $\sst{c}_\O$ value.
Dotted lines show the value of $\sst{c}_\A$ and $\sst{c}_\B$ at unstable fixed-points, with dot-dash lines indicating the corresponding $\sst{c}_\O$ value.
Thick lines mark symmetric fixed points ($\sst{c}_\A=\sst{c}_\B$) and thin lines mark asymmetric fixed-points ($\sst{c}_\A\neq\sst{c}_\B$).
}
\label{fig:bifurcations}
\end{figure}

In Fig.~\ref{fig:bifurcations}, we provide a closer look at the bifurcations of the chemical dynamics in a system with two colloids. 
Depending on the parameter values, we can have either one, (two), three, five, (seven), or nine fixed-points. The bracketed fixed-point counts occur on the onsets of the saddle-node bifurcations, which require exact parameter tuning. 
At $s=0$, we have one stable fixed-point at $\sst{c}_\A=\sst{c}_\B=\sst{c}_\O=0$. Initially, as $s$ is increased we continue with one stable fixed-point. The fixed-point is symmetric ($\sst{c}_\A=\sst{c}_\B$), but for $s>0$ the fixed-point internal concentration values generally are not equal to the external concentration. 
As $s$ is increased further, we reach our first bifurcation, a saddle-node bifurcation which creates two new symmetric fixed-points, one stable and one unstable. Upon increasing $s$ slightly more, we see a pitchfork bifurcation leading to the creation of an asymmetric unstable fixed-point (and its counterpart obtained by swapping $c_\A$ and $c_\B$). Thus, at this point we have two stable and three unstable fixed points.
Increasing $s$ further results in a saddle node bifurcation which creates two new asymmetric fixed points, one stable and one unstable. At this point we are in the regime containing an asymmetric stable fixed-point, which corresponds to non-reciprocal chasing interactions.
Due to the $c_\A \leftrightarrow c_\B$ symmetry of the system, any asymmetric fixed-points have a dual corresponding to swapping $c_\A$ and $c_\B$. Thus this bifurcation is actually two simultaneous saddle-node bifurcations, resulting in a total of nine fixed-points.
As $s$ is increased further, we go through the reverse of these bifurcations, stepping down from nine to five, then to three, and then back down to one fixed-point.

These bifurcations can all be seen in Fig.~\ref{fig:bifurcations}. 
A similar diagram is shown in Fig.~3\p{a} of the main text. However, for the parameters used in the main text, the regions with one and with nine fixed-points dominate, such that three or five fixed-points only appear for very narrow bands of $s$ values. Three or five fixed-points in the chemical dynamics correspond to mutual attraction and mutual repulsion being possible between two colloids in the full spatial system. This can be seen by the thin yellow band in Fig.~3\p{b} of the main text.

\subsection{Finding basins of attraction}
To find the basins of attraction plotted in Fig.~2, we solve the deterministic space-free chemical dynamics using the \texttt{solve\_ivp} function in \texttt{SciPy} \cite{SciPy}, using a Runge-Kutta method \cite{Dormand1980JoCaAM}. We discretize the $(c_\O,c_\A)$ space, simulate the dynamics starting with each $(c_\O,c_\A)$ pair, and colour each point based on which fixed-point the dynamics is attracted to at long times.

\subsection{Solving the stochastic ODEs}
In Fig.~2\p{b}, we plot stochastic trajectories of the chemical dynamics in the presence of chemical noise. 
With chemical noise, the dynamical equations are 
\begin{align}
    \frac{\dif c_\O}{\dif t} &= s K + \Gamma \sum_{i} (c_i-c_\O) - \gamma c_\O \\
    \frac{\dif c_i}{\dif t} &= f(c_i) + \Gamma (c_\O-c_i) + \sqrt{2Q c_i(t)} \,\xi_i(t).
\end{align}
We solve the coupled stochastic ODEs using the \texttt{brainpy} Python package \cite{BrainPy}, using a time-step of $10^{-3}$.

\section{Characterizing the interactions between two colloids}
To produce the diagrams in Fig.~3\p{c} characterizing the two-colloid interactions we use the following procedure.
For each parameter combination, we find the stable fixed-points of the space-free chemical dynamics.
For each stable fixed-point, we initialize a system with the initial chemical concentrations given by the fixed-point solution. The colloids are positioned at a separation $1.2\sigma$.
We then allow the chemical dynamics to evolve with $\mu=0$ up until $t=5$, in order for the chemical dynamics to reach the steady-state values of the spatial system. Then we turn $\mu$ on and inspect the resulting motion of the colloids.
If both colloids move towards or away from each other, then we can characterize the interactions as mutually attractive or repulsive, respectively. Alternatively, if one colloid moves towards the other, whilst the other moves away from the first we classify it as a chasing interaction.


\section{Measuring the winding angle}
For the colloids travelling along a circular confinement, we are interested in the total winding angle $\theta_i(t)$ of each colloid's trajectory around the origin.
To find this, we use the cross product to get the signed increment of the angular coordinate at each time-step
\begin{equation}
    d\theta_i(t) = \hat{\vec{z}}\cdot \frac{\vec{r}_i(t+dt) \times \vec{r}_i(t) }{ |\vec{r}_i(t+dt)| |\vec{r}_i(t)|}.
\end{equation}
Summing the angular increments over time gives the total winding angle
\begin{equation}
    \theta_i(t) = \int_0^t d\theta_i(t).
\end{equation}

\section{Software}
The algorithms for the codes supporting the main findings of this study are available in the paper and its Supplementary Information. Any additional information concerning the code can be made available upon request. 
%
We use \texttt{Matplotlib} \cite{Matplotlib} for plotting and visualisation, in combination with the \texttt{cmasher} \cite{CMasher} and \texttt{cmcrameri} \cite{Crameri2023} colourmaps.

%



\section{Parameter values for figures and supplementary movies}


\textbf{Fig.~2\p{a}}: $N=1$, $a=10$, $b=5.3$, $\gamma=0$. Phase-space flow plotted for $s=\{0.2,0.4,0.8\}$.
\\

\textbf{Fig.~2\p{b}}: $N=1$, $a=11$, $b=5$, $s=0$, $\gamma=0$, $Q=0.2$.
\\

\textbf{Fig.~2\p{c}}: $N=1$, $a=11$, $b=5$, $s=0.3$, $\gamma=0$, $Q=0.3$.
\\

\textbf{Fig.~2\p{d}}: $N=1$, $a=11$, $b=5$, $s=1$, $\gamma=1$, $Q=0.1$.
\\

\textbf{Fig.~2\p{e}}: $N=2$, $a=11.5$, $b=5$, $\gamma=250$.
\\

\textbf{Fig.~3\p{a}}: $N=2$, $a=12.5$, $b=5$, $s=180$, $\gamma=250$, $Q=0$, $\Dc=500$, $\Dp=0$, $\mu=10^5$.
\\

\textbf{Fig.~3\p{b}}: $N=2$, $b=5$, $\gamma=250$, $Q=0$, $\Dc=500$, $\Dp=0$, $\mu=10^5$.
\\

\textbf{Fig.~3\p{c}}: $N=2$, $a=12.5$, $b=5$, $s=180$, $\gamma=250$, $Q=0$, $Q=0.1$, $\Dc=500$, $\Dp=0.05$, $\mu=10^5$.
\\

\textbf{Fig.~3\p{d}}: $N=2$, $a=12.5$, $b=5$, $s=180$, $\gamma=250$, $Q=1$, $\Dc=500$, $\Dp=0.05$, $\mu=10^5$.
\\

\textbf{Fig.~3\p{e}}: $N=2$, $a=12.5$, $b=5$, 
$s(t) = 270-170\sign[\sin(\pi t)]$, $\gamma=250$, $Q=0$, $\Dc=500$, $\Dp=0$, $\mu=10^5$.
\\

\textbf{Fig.~4\p{a}} and \textbf{SuppMov1.mp4}: $N=4$, $a=15.5$, $b=5$, $s=80$, $\gamma=250$, $Q=2$, $\Dc=500$, $\Dp=0.1$, $\mu=10^5$.
\\

\textbf{Fig.~4\p{b}} and \textbf{SuppMov2.mp4}: $N=6$, $\Dc=500$, $\gamma=250$, 
$a=12.5$, $b=5$, $\Dp=0.1$, $Q=2$, $k=10$, $\Rbdr=6$, $\mu=-10^5$,
$s(t) = 330+270\sign[\sin(\pi t/8)]$.
\\

\textbf{Fig.~4\p{c} (top)} and \textbf{SuppMov3.mp4}: $N=2$, $a=12.5$, $b=5$, $s=80$, $\gamma=250$, $Q=0.1$, $\Dc=500$, $\Dp=0.1$, $\mu=-10^5$, $k=10$, $\Rbdr=6$.
\\

\textbf{Fig.~4\p{c} (middle)} and \textbf{SuppMov4.mp4}: $N=6$, $a=12.5$, $b=5$, $s=80$, $\gamma=250$, $Q=0.1$, $\Dc=500$, $\Dp=0.1$, $\mu=-10^5$, $k=10$, $\Rbdr=6$.
\\

\textbf{Fig.~4\p{c} (bottom)} and \textbf{SuppMov5.mp4}: $N=9$, $a=12.5$, $b=5$, $s=80$, $\gamma=250$, $Q=0.1$, $\Dc=500$, $\Dp=0.1$, $\mu=-10^5$, $k=10$, $\Rbdr=6$.


\bibliography{refs_JM}